\documentclass{article}

\usepackage{arxiv}

\usepackage{hyperref} 
\usepackage{float}
\usepackage{amsmath}
\usepackage{amsfonts}
\usepackage{graphicx}
\usepackage{natbib}
 \bibpunct[, ]{(}{)}{,}{a}{}{,}%
\usepackage{multibib}
\newcites{supp}{A1. References}

\newcites{mod}{A2. References}

\usepackage{multirow}

\title{Metawisdom of the Crowd: Experimental Evidence of Crowd Accuracy Through Diverse Choices of Decision Aids}

\author{ \href{https://orcid.org/0000-0000-0000-0000}{\hspace{1mm}Jon Atwell} \\
	South Park Commons\\
	\texttt{jon@finegrained.io}\\
	\And
	\href{https://orcid.org/0000-0000-0000-0000}{\hspace{1mm}Marlon Twyman II} \\
	Annenberg School of Communication\\
    University of Southern California\\
    \texttt{marlontw@usc.edu}\\
}

\date{}

\renewcommand{\shorttitle}{\textit{arXiv}}

\hypersetup{
pdftitle={Decision Aid Choice and Crowd Wisdom},
pdfsubject={q-bio.NC, q-bio.QM},
pdfauthor={Jon Atwell, Marlon Twyman},
pdfkeywords={Wisdom of the Crowd, Collective Intelligence, Decision Making, Decision Aids, Information, Frames},
}

\begin{document}

\maketitle
\begin{abstract}
The provision of information can improve individual judgments but also fail to make group decisions more accurate; if individuals choose to attend to the same information in the same manner, the predictive diversity that enables crowd wisdom may be lost. Decision support systems, from search engines to business intelligence platforms, present individuals with decision aids---relevant information, interpretative frames, or heuristics---to enhance the quality and speed of decision-making but potentially influence judgments through the selective presentation of information and interpretative frames. We describe decision-making as often containing two decisions: the choice of decision aids followed by the primary decision, and define \textit{metawisdom of the crowd} as any pattern by which individuals' choice of aids leads to higher crowd accuracy than equal assignment to the same aids, a comparison that accounts for the information content of the aids. The theoretical model accounting for aid bias and variance shows that an optimal distribution of aid usage can produce metawisdom based on the characteristics of aids within a collection. Three studies---two estimation tasks (N=900, 728) and the nowcasting of inflation (N=1,956; across three collections)---support this claim. Metawisdom emerges from the use of diverse aids, not through widespread use of the aids that induce the most accurate estimates. Thus, the microfoundations of crowd wisdom appear in the first choice, suggesting crowd wisdom can be robust in information choice problems. Given the implications for collective decision-making, the insights warrant future research investigations into the nature and use of decision aids.
\end{abstract}

\section{Introduction}
The possibility of group decisions being more accurate than those of individual experts is a powerful argument for social institutions ranging from representative democracy to markets \citep{aristotle_politics_1967,landemore_democratic_2012,hayek_use_1945,treynor_market_1987,page_difference_2007} and the impetus for efforts to marshal ``crowd wisdom'' for public and private benefit \citep[c.f.][]{wolfers_prediction_2004,dreber_using_2015,cowgill_corporate_2015}.  There are many conditions that give rise to crowd wisdom \citep{larrick_intuitions_2006,davis-stober_when_2014}, but it is generally assumed that crowd accuracy increases when individual deviations from the true value are independent \citep{hong_groups_2004,larrick_intuitions_2006,hertwig_ecological_2019}. Independent deviations ensure errors are distributed, which allows some individuals' errors to cancel each other out. The diversity prediction theorem \citep{hong_groups_2004,page_difference_2007}, colloquially written as \textit{group error = individual error - predictive diversity}, quantifies this cancellation as predictive diversity---the average squared error from the group mean\footnote{The identity is $(C- Y)^2 = \frac{1}{n}\sum_{i}^n(x_i - Y)^2 - \frac{1}{n}\sum_{i}^n( x_i - C )^2$, where $Y$ is the true value, $x_i$ is the prediction of the $i^\text{th}$ crowd member and $C = \frac{1}{n}\sum_{i}^n x_i$.}.

Unfortunately, non-independence of judgments often threatens crowd accuracy. In particular, processes that improve the average individual judgments can increase correlations between predictions and reduce predictive diversity in a crowd. This trade-off means that many group decision-making processes are a balance between decreases in individual error and decreases in predictive diversity. Accuracy increases if predictive diversity decreases more slowly than the average individual error; if not, the crowd becomes less accurate and can indeed make quite poor decisions \citep{janis_groupthink_1982,sunstein_infotopia_2006,caplan_myth_2007}. 

In what follows, we forego the exploration into why individuals make their choices in order to focus on the role of the distribution of aid choices in determining crowd accuracy. To hone the intuition that accuracy can come from a sizable fraction of the group accurately predicting which aid would help them the most, we develop a theoretical model in which usage is proportional to the bias and variance embedded within each aid. The model reveals that proportionally distributing a group can outperform assigning decision-makers in equal proportions to aids and to the single most accurate aid. In our empirical investigations, we observe that groups who choose their own aids can also realize performance gains beyond equal assignment and the single most-accurate aid. These patterns indicate that decision aid choice can offer performance closer to the optimally distributed proportions of aids. 

A thriving area of research considers the effects of peer-to-peer influence processes on group accuracy  \citep{sunstein_making_2014}. Prior research has documented decreases in group errors through information sharing and persuasion \citep{shi_wisdom_2019,becker_wisdom_2019,almaatouq_adaptive_2020} while also documenting increases in group errors through herding behavior \citep{lorenz_how_2011,frey_social_2021} or by suppressing valuable private information \citep{lorenz_how_2011,da_harnessing_2020}. The outcome can also depend on the type of judgment task \citep{guilbeault_probabilistic_2021,becker_crowd_2021}. In the current study, we consider another process through which individual decisions can become correlated; for many scenarios, decision makers attend to a limited set of information cues and cognitive tools to aid their judgments. Following the literature on decision support systems, we refer to these as \emph{decision aids} to emphasize that another actor created or assembled them to improve decisions.

Consider graduate school admissions, a form of ``deliberative bureaucracy'' \citep{posselt_chapter_2016} in which departments pool the judgments of several evaluators to make decisions \citep{michel_graduate_2019}.  To facilitate the evaluation process, departments assemble decision aids related to an applicant's potential, typically standardized test scores, transcripts, letters of recommendation, and personal statements. Ideally, each aid acts as an unbiased signal of potential, and the evaluators aggregate various signals to make accurate decisions. In practice, however, the aids can be biased and contain ``cues that activate the particular construct or scheme in question---are a part of the to-be-evaluated information itself'' \citep[p. 4]{gilovich_where_2012}. This influences how the evaluators process the information the aids contain\footnote{Standardized test scores can make mathematical achievement more salient than warranted \citep{koretz_measuring_2008}, transcripts can frame coursework around grades in a way that can penalize ambitious or unconventional choices \citep{berry_individual_2009}, the evaluation of personal statements tends to reflect assumptions about fit \citep{chiu_its_2019}, and letters and research experience can privilege the well-connected or financially well-resourced \citep{douglass_undergraduate_2013}.} \citep{kahneman_judgment_1982,gilovich_judgment_2010,keren_wiley-blackwell_2015}.

Decision support systems exist to create a decision-making process in which ``relevant knowledge will be brought to bear at the point where the decision is made'' \citep[p.82]{simon_administrative_1957}. We argue this intention is often undermined by users' preferences for single indicators or heuristics \citep{gigerenzer_simple_1999,zsambok_naturalistic_1997} and limited attention \citep{bordalo_salience_2022}. The more decision aids inform individual judgments, the more important the pattern of aid usage across a group becomes in determining the accuracy of the group. One pattern may threaten group accuracy; if evaluators rely on the same subset of decision aids, their predictions will likely be correlated, and vital predictive diversity in the group will be lost. In the case of admissions, evaluators might rely only on standardized test scores---a consistent worry in admissions \citep{koretz_measuring_2008,posselt_chapter_2016}---because scores are the single most predictive aid and are easy to digest. Doing this is reasonable for individual judgments, but this pattern could lead to worse group decisions as relevant information is left unused\footnote{In fact, the combination of scores and undergraduate grades can predict better than scores alone \citep{kuncel_standardized_2007}.}.

On the other hand, another pattern presents an opportunity. Groups can be collectively accurate \textit{because} they make diverse choices about which aids to use. That is, crowd wisdom can be traced to the fact that across two judgments (the first being which aids to use, and then the next is the primary judgment), individuals often make suitably diverse predictions. We argue this pattern is likely. If individuals want to make accurate primary judgments, they have an implicit goal of choosing decision aids that reduce their own error, whether through the recognition of what information is valuable for making the judgment, metacognitive matching of individual cognitive strengths to information that supports their strengths \citep{flavell_metacognition_1979}, or even the act of deciding, which can reduce errors by encouraging reflection \citep{keren_users_2015}.  Crucially, when \textit{only some} individuals are accurate in this first decision, the distribution of primary predictions will contain essential predictive diversity because of the erroneousness of others.

To assess the degree to which crowds might realize such potential, we conduct an exploratory study (N=900) and two preregistered studies (N=1,956 and N=728). Each study distinguishes the effect of aid choice on accuracy contrasted with the effect of the information present in a set of decision aids by comparing a crowd choosing aids to a crowd in which aids are randomly assigned in equal proportions. In the exploratory study, participants estimated the number of candies in a bowl with the help of one of three aids. Aid use was not distributed inversely proportional to error, but there is predictive diversity, and the crowd choosing their aids outperforms the one assigned to the aids. The first preregistered study has crowds forecast the Consumer Price Index (CPI) on three different occasions. For two of the three monthly figures, the choice of aid outperforms aid assignment. For the third, all three aids erred in the same direction, a fact the model suggests reduces the relative performance gain of choice. In the second preregistered study, individuals rank their aid choices and provide a distributional estimate for the number of coins in a glass cylinder. Again, choice outperforms assignment, and we show this is uniquely attributable to individuals' choice of aid. Cumulatively, the empirical investigations offer insights that demonstrate the applicability of the proposed theoretical model for explaining how decision aids impact the accuracy of a crowd.

We conclude that the use of decision aids can improve group decision-making not only by providing relevant information but also by inducing predictive diversity. Crucially, individuals' choices about which aids to use effectively calibrate the trade-off between improving individual judgments and preserving predictive diversity. This finding suggests the need for more research into the composition of decision aids within decision support systems, the origins of individual choices, and the further effects of deliberation and influence.

\subsection{The Wisdom of the Crowd and Decision Support Systems}
To introduce our perspective on the origins of crowd wisdom, we first describe the wisdom of crowd phenomenon in terms of two microfoundations and two broad types of mechanisms \citep{landemore_collective_2012-1}. All crowds can be assessed based on the microfoundations of individual ability and predictive diversity. Predictive diversity---the average error from the group mean---is a common focus in contemporary research, but individual ability can be the overriding factor. Under some conditions, individual ability is the core microfoundation because more predictive diversity means individuals are more likely to be wrong. However, research has focused on how the quantitative relationship changes between individual ability and predictive diversity in the context of different mechanisms for producing the crowd's judgment \citep{berend_when_1998}.

When individual accuracy and predictive diversity are present, actual crowd judgments arise through two broad types of mechanisms: deliberation and institutionalized aggregation procedures such as majority voting, averaging, or markets \citep{landemore_collective_2012-1}. These two broad types of mechanisms can work in sequence---as with the Delphi method of eliciting private judgments, then deliberating, and finally voting privately---but a recent stream of research focuses on the more entwined and often network-based processes in which aggregation happens in the course of peer-to-peer deliberation and social influence. The fact that outcomes can differ substantially highlights the importance of both microfoundations and precise mechanisms \citep{mavrodiev_ambigous_2021,yan_paradox_2021}. 

Social influence undermines group accuracy in many circumstances \citep{lorenz_how_2011, lorenz_majoritarian_2015, frey_social_2021}, but also can improve accuracy when it occurs in decentralized networks \citep{jonsson_kind_2015,becker_network_2017} when participants adapt their ego network as they receive information \citep{almaatouq_adaptive_2020}, or when a larger group is partitioned into smaller subgroups \citep{navajas_aggregated_2018}. Additionally, different task types or voting procedures reveal the same sensitivity \citep{minson_contingent_2018,prelec_solution_2017,marshall_quorums_2019,becker_crowd_2021}. Also, markets can sometimes accurately predict outcomes \citep{wolfers_prediction_2004,servan-schreiber_prediction_2012,cowgill_corporate_2015,peeters_testing_2018}, but are also susceptible to extreme inaccuracy when information is sparse \citep[ch. 4]{sunstein_infotopia_2006} or uncertainty is high \citep{rulke_herding_2016}.  In the current study, we accept that crowds will reflect many of these external aspects based on environmental and social factors\footnote{For some exceptions, see \cite{mellers_psychological_2014,economo_social_2016,bennett_making_2018,keck_enhancing_2020,da_harnessing_2020}.}, and we also acknowledge that individual accuracy and predictive diversity are a direct consequence of the collection of actors assembled. Increasing the weight given to the inputs of experts should increase average group accuracy \citep{lee_inferring_2012,mannes_wisdom_2014,budescu_identifying_2015} and adding individuals with diverse experiences and aptitudes---generally labeled cognitive diversity---should increase predictive diversity \citep{page_difference_2007,davis-stober_composition_2015} \footnote{Cognitive diversity might be associated with demographic diversity but need not be \citep{de_oliveira_demographically_2018}.}.

Individuals' searches for information and interpretations are vital sources of cognitive diversity in groups, but decision support systems supplant that search with a curated set of facts and frames for interpreting them. Decision support systems have been widespread for several decades \citep{keen_decision_1987,ain_two_2019} and encompass communication support, process structuring, and information processing elements\footnote{Because of its generality and descriptive clarity, we use the term decision support systems to encompass a wide range of approaches and alternative names (e.g., expert systems, executive information systems, business intelligence systems, group decision support systems, dashboards, key performance indicators, and others).} \citep{desanctis_foundation_1987,humphreys_user-centered_1996,zigurs_theory_1998}. The designers of these systems ``have often concentrated on `enhancing' the judgment environments by providing decision aids and interventions designed to make judgment more accurate, and in doing so have made assumptions about what individuals need to know and how this information should be displayed'' \citep[p. 80]{mosier_technology_2008}.

While striving to meet the goal of bringing relevant information to ``the point where the decision is made'' \citep[p.82]{simon_administrative_1957}, these systems can be imperfect.
Business intelligence systems implement dashboards or scorecards of key performance metrics to guide teams' decisions \citep{rigby_management_2011,allio_strategic_2012}, but the chosen metrics can omit important information \citep{kaplan_balanced_1992}, visual presentations can bias judgments in subtle ways \citep{bremser_developing_2013}, and users' attention is often intentionally drawn to specific metrics \citep{few_information_2013}. For example, retail investors often choose investments based on an extreme subset of easily accessible performance metrics, but different presentations of information and choices can affect individuals' choice of investments \citep{benartzi_risk_1999,benartzi_naive_2001,diacon_framing_2007}. Similarly, nonpartisan voter guides are supposed to inform voters \citep{mummolo_how_2017,boudreau_roadmaps_2019}, but frames that emphasize ``a subset of potentially relevant considerations'' \citep[p. 230]{druckman_implications_2001} have been found to alter political behaviors by drawing focus to particular aspects of public discourse \citep{nelson_issue_1999,scheufele_framing_1999,diacon_framing_2007}.

Regardless of the context for a decision support system, it creates a two-stage decision-making process. Before rendering the primary judgment, decision support systems put users in the position of choosing what information to utilize. This choice is a decision where a crowd can exhibit wisdom; individuals can make good decisions about what information or interpretative tools will be helpful, and the decisions on this initial task may be a source of both individual accuracy and predictive diversity. Thus, a fundamental question for the crowd wisdom literature is whether a group's collective choices for the first decision reduce individual error more than they remove predictive diversity or vice versa. Do group members choose aids with more or less biasing potential? Relatedly, do they choose aids such that the final judgments ``bracket'' the true value, allowing different errors to cancel each other out? To begin to address these questions, we formalize two distinct effects for decision aid use and decision aid choice.

\subsection{Decision Aid Use in Crowds}

We represent the value being estimated as $Y$, a random variable of finite mean and variance. Let random variable $X_i$ be the prediction of decision-maker $H_i$. The \emph{crowd estimate} is then $C=\sum_i^N p_iX_i$, where $p_i$ is the proportion of the sum given to the estimate of $H_i$, $X_i$. The distribution of individual errors is captured by the \textit{mean squared error} (\textit{MSE}) or $\frac{1}{N}\sum_i^N(X_i - Y)^2$. The \textit{group squared error} (\textit{GSE}) is $(C-Y)^2$ and is the commonly used measure of group accuracy. Following the generalized definition of crowd wisdom from \citet{davis-stober_when_2014}, a crowd is wise anytime the \textit{GSE} is less than or equal to the \textit{MSE}. By Jensen's inequality, this is always the case for simple averages (i.e., $p_i = \frac{1}{N})$, so we focus on the relative accuracy of different crowds. 

Let $\mathcal{D}$ to be the set of available decision aids, and let $S_j$ be any nonempty subset of $\mathcal{D}$. We assume that $X_i$ is conditional on the particular $S_j$ decision-maker $H_i$ is exposed to. The hypothetical difference $E[X_i |S_j] - E[X_i | \emptyset]$, where $\emptyset$ is the empty set, is the treatment effect of decision aid set $S_j$ on $H_i$. Because this difference is unobservable, we use the average treatment effect under random assignment.

We define the information effect of $\mathcal{D}$ as the average treatment effect for a treatment group receiving no aids and a treatment group with randomized assignments to each of the individual decision aids in $\mathcal{D}$. With $C^{\emptyset}$ as the crowd estimate of a population of $H_i$ that receives no decision aid treatment and $C^\text{random}$ as the crowd estimate of a population that receives randomized exposures to individual aids, the information effect of $\mathcal{D}$, $IE_{\mathcal{D}}$, is the following:
\begin{equation}\label{eq:info_effect}
IE_\mathcal{D} = E[(C^{\emptyset} - Y)^2] - E[(C^\text{random} - Y)^2]
\end{equation}
There is an information effect for $\mathcal{D}$ whenever randomized exposure to the aids of $\mathcal{D}$ results in nonzero difference. When the second term is smaller, the information treatment is positive and represents an improvement in crowd accuracy through aid use. A negative information effect is possible if $\mathcal{D}$ contains low-quality decision aids that induce errors in the same direction on average, likely increasing individual errors faster than diversity is introduced. 

Finally, to establish the role of active choice for a particular $S_j$ in altering crowd accuracy, we define the \emph{choice} effect of $\mathcal{D}$, $CE_{\mathcal{D}}$, as the difference in crowd errors between $C^\text{random}$ and a crowd choosing which aids to view, $C^\text{choice}$.
\begin{equation}\label{eq:meta}
CE_\mathcal{D} = E[(C^\text{random} - Y)^2] - E[(C^\text{choice} - Y)^2]
\end{equation}
Any scenario in which $CE_\mathcal{D}$ is positive is a phenomenon we term the \emph{metawisdom of the crowd}. That is, the crowd exhibited wisdom in the first stage that leads to greater accuracy in the second, primary task above and beyond the effect of exposure to additional information. 

Whereas $IE_\mathcal{D}$ is an average treatment effect, $CE_\mathcal{D}$ is not because we can no longer assume a lack of correlation between treatment group and distributions of estimates; the introduction of correlations through metacognitive processes is one of the justifications for looking at the effect of choice. This implies that the mean estimate associated with having chosen to view a given aid set can be different from the mean estimate under random assignment.

The availability of decision aids and choice among them is a crucial lens through which to understand the origins of crowd wisdom. Reflecting the goal of decision support systems, we assume there is a positive information effect for a well-constructed set of aids (H1). 
\begin{equation}\label{eq:hypo1}
H1: E[(C^\text{random} - Y)^2] \leq E[(C^{\emptyset} - Y)^2]
\end{equation}
We also hypothesize that crowds will exhibit metawisdom, a positive effect of choice (H2). This hypothesis is motivated by the intuition that a group of decision makers will tend to favor aids that improve individual accuracy, but enough individuals will choose aids that preserve vital predictive diversity. We now formalize and explore the intuition with a formal model.
\begin{equation}\label{eq:hypo2}
H2: E[(C^\text{choice} - Y)^2] \leq E[(C^\text{random} - Y)^2]
\end{equation}

\section{Optimal Distribution of Aid Use in Crowds}

Hypothesis testing assumes that there will be circumstances where aid use will be distributed differently among a crowd of decision-makers. For example, a set of aids may be allocated with equal probabilities to all crowd members, or aids may be selected with different proportions of the crowd accessing various aids. Given that group accuracy is a balance between individual accuracy and predictive diversity, there is a competing demand for aid use that maximizes individual accuracy by minimizing error while also preserving diversity through the use of different aids--assuming aids produce distinct predictions. In the presented model, a crowd of decision-makers is analogous to the set of decision aids distributed among them since we assume that individual accuracy is represented by the bias inherent to a decision aid along with the variance of potential estimates stemming from its use.

This assumption offers an explicit explanation for predictive diversity being the result of a set of aids where each aid promotes different estimations from other aids, as well as potentially within the aid. To represent this assumption, we consider the bias-variance tradeoff for aid use. Recall that the definition of the crowd estimate is a linear combination
\begin{equation}
C=\sum_i^N p_iX_i
\end{equation}
where $p_i$ is the weight associated with the prediction $X_i$ made by a decision-maker using a specific decision aid, and $\sum_{i=1}^N p_i = 1$. In the model, we consider the prediction $X_i$ to be influenced by the characteristics of a decision aid. For each $X_i$, we assume an aid has a constant associated bias ($B_i)$ and variance ($\sigma_i^2$) that are observable surrounding the true mean ($\mu$) for a crowd's estimate of $Y$ (the value being estimated) when using the aid. As an assumption,  $\varepsilon_i$ is the random noise term for the prediction: 
\begin{equation}
X_i = \mu + B_i + \varepsilon_i
\end{equation}

\[\mathbb{E}[\varepsilon_i] = 0 \text{ and } \text{Var}(\varepsilon_i) = \sigma_i^2\]

Thus, \[B_i = \mathbb{E}[X_i] - \mu\]

Therefore, \[C - \mu = \sum_{i=1}^N p_i (X_i - \mu) =\sum_{i=1}^N p_i \bigl(B_i + \varepsilon_i\bigr)\]

Under such conditions, we can then minimize the group squared error of $C$, i.e., \[\mathrm{GSE}(C) = \mathbb{E}\bigl[(C - \mu)^2\bigr] \]

Now, if we assume $\varepsilon_i$ of the aids are uncorrelated, then
\begin{equation}
\mathrm{GSE}(C)=
\mathbb{E}\Bigl[\bigl(\sum_{i=1}^N p_i B_i + \sum_{i=1}^N p_i \varepsilon_i\bigr)^2
\Bigr]
=\underbrace{\Bigl(\sum_{i=1}^N p_i B_i\Bigr)^2}_{\text{square of pooled bias}} + \underbrace{\sum_{i=1}^N p_iB_i\sum_{i=1}^N p_i\varepsilon_i}_{\text{cross-terms}} +  \underbrace{\sum_{i=1}^N p_i^2 \varepsilon_i^2}_{\text{pooled variances}}
\end{equation}
The cross-terms are eliminated because \[\mathbb{E}[\varepsilon_i]=0 \text{ and } \mathrm{Cov}(\varepsilon_i,\varepsilon_j)=0 \text{ for } i\neq j\]
Hence, \[\mathrm{GSE}(C)=\Bigl(\sum_{i=1}^N p_i B_i\Bigr)^2
 + \sum_{i=1}^N p_i^2 \sigma_i^2\]
The decomposition of the group squared error based on bias and variance allows for the assessment of optimality in terms of two critical features of an aid that directly impact individual accuracy and predictive diversity.

\subsection{Optimal Proportions of Aids to Minimize Error}

Given this description of how aid use influences group estimation, the optimal distribution of decision aids among crowd members would be when aids are distributed in a manner inversely proportional to the errors of the aids themselves. In such a case, aids that result in estimates with smaller errors should be selected more frequently than those that produce larger errors. We employed the Lagrange Multipliers technique to determine the specific weightings that will minimize error (see Supplemental Materials Appendix \ref{sm: calculate} for derivation). The optimal aid proportions can be expressed as

\begin{equation}
    p_i = \frac{1}{\sigma_i^2} \cdot \frac{\left(1 + \sum_{k=1}^{n} \frac{B_k^2}{\sigma_k^2}\right) - B_i \sum_{k=1}^{n} \frac{B_k}{\sigma_k^2}}{\left(\sum_{k=1}^{n} \frac{1}{\sigma_k^2}\right) \left(1 + \sum_{k=1}^{n} \frac{B_k^2}{\sigma_k^2}\right) - \left(\sum_{k=1}^{n} \frac{B_k}{\sigma_k^2}\right)^2}
    \label{eq:optimal_aid_proportion}
\end{equation}

For a set of aids, different proportions are assumed to sum to 1, which constrains the proportion derivation. Based on the summations, there are indications that both aid bias and variance greatly contribute to the optimal proportion of a given aid. For simplicity, considering the proportional representation of the optimal solution is productive when addressing specific cases of aid distribution: 

\begin{equation}
    p_i \propto \frac{\left(1 + \sum_{k=1}^{n} \frac{B_k^2}{\sigma_k^2}\right) - B_i \sum_{k=1}^{n} \frac{B_k}{\sigma_k^2}}{\sigma_i^2}
    \label{eq:proportional}
\end{equation}

From the derivation, and to account for scenarios most relevant to our subsequent empirical investigations, three cases must be considered in addition to the relatively complex general case above: when estimation variances within all aids are zero (only considering bias), all aid biases are zero (only considering variance), and when estimation variances across aids are equal (exploring bias-variance tradeoff). The cases are necessary because the generalized optimal distribution encapsulates multiple scenarios that affect the overall estimation capabilities of the crowd. With these three cases, we can illustrate multiple considerations one should make regarding aid selection to provide an intuition for how the underlying qualities of aids impact the ideal distribution of aids--either through assignment or choice. The core theoretical arguments surrounding our developed concept of metawisdom arise directly from our discussion of these cases.


The first case is concerned only with the biases of aids. Specifically, we assume that each decision-maker produces the same estimate depending only on the bias within a given aid; all error is derived from the bias embedded within an aid. When variance is zero, minimizing the group squared error relies only on bias, and we can use a simplified calculation of $\mathrm{GSE}(C)=\Bigl(\sum_{i=1}^N p_i B_i\Bigr)^2$. With this setting, any minimization will be determined based on the nature of the aids themselves. 

In one scenario, if a set of aids includes differently-signed biases (over- or underestimation) of various magnitudes, it then becomes possible to calculate proportions that will allow for error cancellation among aids and the overall error to approach zero. Relatedly, available aids need not be selected if they do not contribute requisite error reduction or predictive diversity. Another scenario to consider is when all the biases are in the same direction; in this scenario, the entire crowd selecting the least biased aid that minimizes estimation error will then be considered the optimal distribution, despite an expected reduction in prediction diversity. More generally, there are many potential optimal distributions of aid use proportions depending on the attributes of aids and how aids compare to other aids in the set. We further investigate these dynamics by implementing a computational model to illustrate how aid biases interact with one another and result in crowds achieving different accuracy levels during estimation (see Supplemental Materials Appendix \ref{sm: computation}). Different aid biases allow for the emergence of metawisdom when the aid set is structured in a manner where predictive diversity may be integrated into the group-level estimation. Metawisdom then becomes a helpful term to describe situations when a group of people distribute themselves proportionally into aid use patterns that minimize error by reducing the impact of biases that some crowd members are exposed to through their selections.

For the next case, we consider situations where we have unbiased aids, but they still contain estimation variance. If $B_i=0$ for all $i$, then $\mathrm{GSE}(C) = \sum_{i=1}^N p_i^2 \sigma_i^2$ and the optimal aid proportions are distributed inversely proportional to the variances, $p_i \propto 1/\sigma_i^2$. In this situation, a set of aids may be differentiated from one another by evaluating the group accuracy for each aid, and justification for the allotment of multiple aids stems from attempting to manage the variance that emerges from usage. For example, an aid with high variance may be deemed necessary for the promotion of predictive diversity even when lower variance aids are available. For such situations, it is also worth noting that the absence of aid bias would suggest that selecting only the low variance aids corresponds to lower error, but assessing the predictive diversity from estimations based on a set of aids may not reveal appreciable differences in estimation distributions. Specifically, there are cases where aids with relatively higher and lower variances yield similar distributions and less sensitivity based on the task under investigation. When considering only variance, more flexibility is needed when determining strategies to minimize estimation error while still maintaining predictive diversity.  

Finally, we return to the more general case that reflects a bias-variance tradeoff for aid usage. There is an inherent complexity in the problem formulation stemming from the nature of how aid bias is coupled to the proportion of a crowd exposed to an aid. Generally, $p_i \propto 1/(\sigma_i^2 + B_i^2)$ does not necessarily hold when $B_i \neq 0$, because the $\bigl(\sum_i p_i B_i\bigr)^2$ term depends \textit{linearly} on the $p_i$ before squaring to illustrate the relationship between the embedded nature of bias within a particular aid. Meanwhile, the $\sum_i p_i^2 \sigma_i^2$ term depends on the \textit{squares} of the $p_i$ describing a different rate of influence arising from the variance term. As a result, identifying the optimal proportion of an aid (i.e., Eq. \ref{eq:optimal_aid_proportion} and Eq. \ref{eq:proportional}) is contingent upon relationships among the entire set and requires empirical investigation to assess how aid use dynamics may shift in response to specific aids as well as scenarios. These make identifying patterns of metawisdom a function of aid characteristics, individual estimation, and distribution of aids throughout a group; a multilevel set of concerns. This accepted complexity directly guides our reported empirical investigations where we focus attention on whether individual choices in aid use can offer improvements or at least differences from more homogenous aid distributions. 

\section{Decision Aid Experiments}
We conducted three IRB-approved decision aid choice experiments with English-speaking adults based in the United States. Participants were recruited from the Prolific research panel\footnote{\url{https://www.prolific.com}} and provided informed consent before beginning the tasks. They were compensated for task completion and the accuracy of their predictions. Each of the studies contained two conditions: \textit{Assigned}, in which participants are randomly assigned to a decision aid as a part of the estimation task, and \textit{Choice}, in which participants select a single aid using stylized visual and textual descriptions of the available aids before making an estimate. The first two studies also contained a \textit{No Aid} condition, wherein participants responded to a prompt without any decision aid. For the third study, all participants provided an initial, unaided estimate instead. 

\subsection{Study 1: Candy Count Estimation Task}
The first experiment is a quantity estimation task similar to the classic ``Treynor'' task in which participants estimate the number of beans in a clear glass jar \citep{jenness_social_1932,treynor_market_1987}. It serves as an initial exploration of how choosing a decision aid impacts a well-known and tightly controlled ``wisdom of the crowd'' task. 900 participants earned \$0.65 for completing the task. Participants also received a \$2 bonus for estimates within 5\% of the true value. The most accurate estimate received \$25; if there had been a tie, all winning guesses would have split a \$50 pot. Participants confirmed understanding of this compensation structure. 

The task asked participants to estimate the number of multicolored candies in a glass bowl viewed from an angle that showed both the top and sides of the bowl. Before submitting an estimate, participants viewed one of three aids. When choosing among the three aids, participants viewed a stylized visual representation and text description for each aid. To address the possibility of an ordering effect, the order of choices was randomized. The decision aids all show the primary image and were designed to invoke both explicit and implicit reasoning steps, but to varying degrees. One decision aid gives the volume of the bowl and the mean volume of an individual candy piece. This \emph{equation} aid requires explicit reasoning about the volumes, but participants must also estimate the packing coefficient (the ratio of candy to air in the container). In contrast, the \textit{comparison} aid promotes reliance on more intuitive reasoning; it shows the bowl from the primary image emptied of candies next to a larger bowl full of the same type of candy. The precise count of candies in the latter bowl is given to explicitly establish an upper bound for estimation, but otherwise requires an intuitive estimation based on the relative bowl sizes. The final aid is the \emph{scale} aid, which adds a ruler to the primary image to calibrate the relative scale of a single candy piece to the bowl. Those using the aid could explicitly reason about the count, but the number of necessary calculation steps likely causes users to rely on more intuitive reasoning. No other aids were designed or tested. Additional notes about task design elements, including a condition in which participants can view multiple aids, are available in the Supplemental Materials Appendix \ref{sm: study1}.

\subsubsection{Study 1 Results}
\begin{figure}
\centering
\includegraphics[scale=.1]{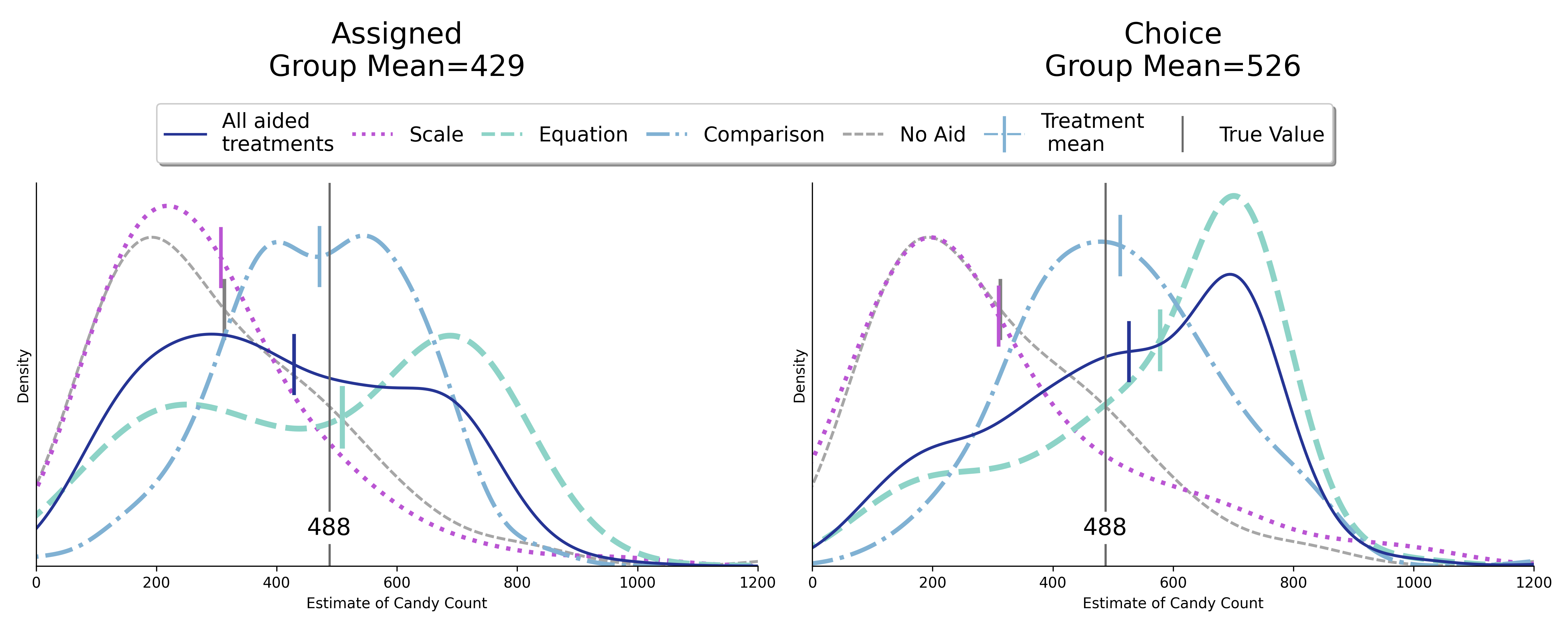} 
\caption{Study 1: Candy Count Estimation Task. The kernel density estimates (KDE) of the crowd estimation distributions for the Assigned (left) and Choice (right) conditions. The vertical line is the true value, 488 candies, and the vertical hashes on the KDEs mark the average crowd estimate for participants using the corresponding decision aid.}
\label{fig:bean_choice}
\end{figure}

Figure \ref{fig:bean_choice} visualizes the results of the experiment using kernel density estimates (KDEs) of the distributions of crowd estimates for the conditions. The left panel contains the Assigned condition, and the right contains the Choice condition. The No Aid condition is plotted in both panels. The vertical hash on each KDE is the crowd estimate for that aid. The vertical black line shows the true count of candies (488). The shapes of the KDEs suggest an effect of decision aid use, and the Kolmogorov-Smirnov test ($p<0.001$) confirms significant effects for all three distributions of aided estimates. All differences in means between decision aids are statistically significant ($p < 0.001$; Welsh's T-test with unequal variances) except for the \emph{equation} and \emph{comparison} aids in the Assigned condition ($p=0.185$). 

These differences create the possibility that there will be an information effect for the set of aids. While the mean for \emph{scale} aid is not significantly different from that of the No Aid condition, the crowd estimate across all aids is significantly different from the No Aid estimate ($p < 0.001$, Welsh's T). This confirms Hypothesis 1, the existence of an information effect for the set of aids. There is also a choice effect for the aids, meaning the Choice condition is more accurate than the Assigned condition ($p <0.001$ for each comparison, Welsh's T). This confirms Hypothesis 2, the existence of metawisdom of the crowd.

\begin{table}[h]
\caption{Candy Count Estimation Task: Detailed breakdown of results. $N$ is the number of participants in the respective treatment.}
\label{tab:Table1}
\centering
\small
\renewcommand{\arraystretch}{1.2}
\begin{tabular}[]{ @{\extracolsep{.1em} } c | c || c | c | c | c }
Choice & Decision Aid & N & Mean & Group Squared Error & Mean Squared Error\\
Treatment & Treatment & & Estimate & (GSE) & (MSE) \\

\hline
\hline

\multirow{4}{*}
{Assigned } & {Scale} & 133 & 307 & 32584 & 82599\\
&{Equation} & 133 & 510 & 463 & 87076  \\
&{Comparison} & 134 & 471 & 283 & 23578\\
\cline{2-6}
&{\textit{All}} & 400 & 430 & 3422 & 64316\\

\cline{1-6}

& {Scale} & 44 & 311 & 31353 & 78492\\
{Choice} &{Equation} & 222 & 578 & 8151 & 63771\\
&{Comparison} & 134 & 512 & 600 & 30693\\
\cline{2-6}
&{\textit{All}} & 400 & 527 & 1508 & 54309\\

\cline{1-6}

\hline
\hline
Control & {No Aid} & 100 & 313 & 30685 & 71537\\

\end{tabular}
\end{table}
Table \ref{tab:Table1} details how metawisdom emerged; 11\% of participants in the Choice condition chose the \emph{scale} aid (least accurate), 55.5\% selected the \emph{equation} aid, and 33.5\% chose the \emph{comparison} aid (most accurate). This choice distribution does not correspond to an optimal distribution of aid usage based on inverse-proportionality to error size, yet the distribution of choices preserves predictive diversity; if all participants had chosen the \emph{equation} and \emph{comparison} aids instead of the most inaccurate aid, \emph{scale}, and the effects on errors of the former two aids under choice stayed the same, the overall crowd accuracy would have declined\footnote{Distributing the 44 participants who chose the \emph{scale} aid evenly to the other two aids results in a group estimate of $552 = \frac{244}{400}(578)+\frac{156}{400}(512)$ and group squared error of 4096. Even if all 44 went to the most accurate aid, \emph{comparison}, the group estimate would be 548, and the group squared error (3676) would still be larger than the observed error of the Choice condition.}. These patterns illustrate how the presence of a differently-signed decision aid within a set contributes to enhanced group-level accuracy, even when it is selected by a relatively small minority (11\%) of the group.

Interestingly, we observe metawisdom despite the fact that there are negative choice effects for the individual decision aids, a case of Simpson's paradox. In the Choice condition, the crowd error for \emph{equation} and \emph{comparison}---the two most used aids---increases relative to the Assigned condition. However, the overall group accuracy improved because many participants chose the decision aids that are associated with more accurate estimates. While further research into the cognitive mechanisms of choice is warranted, these dynamics suggest that metawisdom in this case does not come from crowd members choosing an aid that would most help them individually. Rather, participants tend to select what is ultimately an aid with a strong positive information effect, but some proportion of the crowd chose an aid with no information effect and preserved necessary diversity for the crowd.

\subsection{Study 2: CPI Forecasting Experiments}
The task in Study 1 demonstrates an instance of metawisdom of the crowd in a task where individuals were not expected to have prior beliefs or private information. Study 2 is a pre-registered\footnote{\url{https://osf.io/qm7cs/?view_only=7a51d1a3ecb54e9080b40ee946c80bfa}} task about forecasting the consumer price index (CPI), a topic about which participants likely have prior beliefs and private information. Inflation was a salient political issue during data collection, and participants may have had knowledge of trends in the index, if not the most recent figure itself. This knowledge could dampen the effects of decision aids, and this task makes for a more conservative test for the existence of metawisdom. 

Inflation forecasting is also a setting in which crowd wisdom is known to exist. For example, the average expectation for the inflation rate in 12 months' time based on a random sample of United States households does well in predicting inflation rate changes despite households' limited macroeconomic knowledge and focus on individual circumstances \citep{curtin_consumer_2019, verbrugge_whose_2021}. In fact, households predicted the inflation surge of 2021-2022 \citep{sablik_forecasting_2021,glover_disagreement_2022} while central bankers and other experts downplayed the possibility, demonstrating the power of aggregating diverse knowledge and judgments.

The current study task asks participants to forecast the CPI value for the previous month two days before the value and data are publicly released\footnote{For example, we collected predictions on December 8th and 9th, 2021 before the November 2021 figure was released on December 10th by the U.S. Bureau of Labor Statistics.}. Professionals who predict the figure after the relevant period but before the data is released call it \textit{nowcasting}. We repeated this task with different participants, resulting in three distinct data samples across data collection periods for November 2021, May 2023, and August 2023. The replications test the robustness of information and choice effects. Recent changes in the CPI value, including trends and broader economic conditions, have the potential to change the informativeness of the aids and potentially the pattern of aid choices.

We used the same types of aids for all three collections, although the contents of each aid changed. One aid was the graphical depiction of a predictive model for the next 12 months based on the previous 48 months of CPI data (\textit{Predictive Model}). The second aid was the national summary from the Federal Reserve's ``Beige Book;'' the summary is a three-paragraph overview of economic trends and includes a paragraph titled ``Prices'' that provides insights into price trends (\textit{Fed Statement}). The third aid displays the trends of the major subcomponents of the CPI (e.g.,food, energy, and other commodities) for the previous 24 months (\textit{Components}). We designed the third aid to allow participants to more concretely connect their personal experiences to the different price trends, and potentially project from their recent experience to predict the latest figure. All three aids also included a plot of the CPI for the previous 24 months to ensure that the task was more about predicting the change in inflation rather than a measure of prior knowledge about the general range and past trends in the figure \footnote{The Federal Reserve statement aid was a verbatim copy of the text in the Beige Book. The Predictive Model decision aid was created using a calibrated ARIMA model. The Components decision aid was created using data from the website for the Bureau of Labor Statistics.}. No other aids were designed or tested. Submissions were timed to mitigate the use of other resources (90 seconds for \emph{Predictive Model} and \emph{Components}, 120 seconds for \emph{Fed Statement}).

Across the three collections, we recruited a total of 1,956 participants. Each earned \$0.65 for completing the task, estimates within 0.15 of the true value earned a \$2 bonus, and the closest guess(es) split a \$50 reward. Once the true CPI value was announced, participants received a message through the Prolific platform with the value. The prompt to choose an aid provided text descriptions of the aids. Because there was a possible ordering effect on choice based on which aid descriptions were presented, the design was counterbalanced evenly across all six possible configurations. An analysis of the counterbalanced subgroups revealed no significant differences in choices. See Supplemental Materials Appendix \ref{sm: study2} for examples of experimental stimuli.

\subsubsection{Study 2 Results}
\begin{figure}
\centering
\includegraphics[scale=.4]{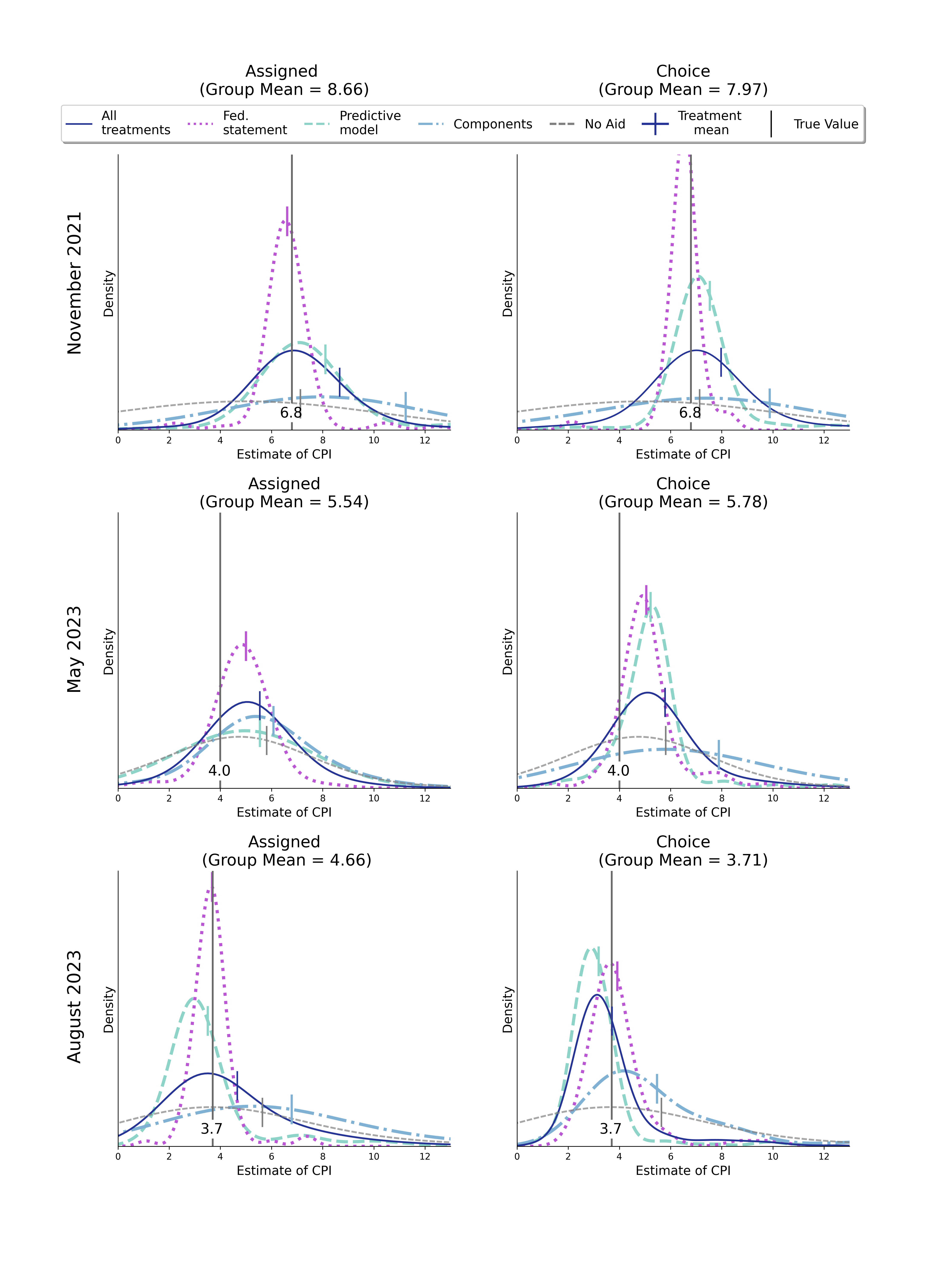} 
\caption{CPI forecasting experiment. The panels contain the kernel density estimates (KDE) for the distribution of predictions by aid type. The left column is for the Assigned condition, and the right column is the Choice condition.  The vertical line in each panel is the true CPI value, and the vertical hashes on the KDEs denote the mean estimate for that treatment condition.}
\label{fig:choice}
\end{figure}

Figure \ref{fig:choice} plots the KDEs for the distribution of predictions by choice conditions and aid type. The left column contains the Assigned conditions, and the right column contains the Choice conditions. Each row corresponds to the different collection periods. The KDEs reveal the decision aids have unique and fairly consistent effects on the distribution estimates across all three collection periods. The \emph{Fed Statement} decision aids produce the most accurate group estimates and tend to have lower variance. The \emph{Predictive Model} aids are the second most accurate, and estimates with the \emph{Components} aids are the least accurate while also having high variance. Interestingly, the No Aid conditions have reasonably accurate group means, although the variance is notably high. This pattern suggests participants may be informed to a certain degree, either through general awareness or contemporaneous searches online. 

These different distributions make the setting ripe for metawisdom of the crowd, our primary preregistered hypothesis (H2). For the November 2021 and August 2023 collections, we observe metawisdom. For the November 2021 collection, this is marginally significant at $p=.058$ (Welsh's T). For August 2023, the effect is significant at $p<0.01$ (Welsh's T). There is no metawisdom for May 2023. In fact, the mean of the Assigned group is more accurate than that of the Choice group. However, our model offers an explanation for these patterns; the lack of a choice effect was likely due to the errors for the individual aids all having the same sign, which prevents error cancellation to the extent needed to outperform the aid displaying the least amount of bias. In such cases, instead of introducing predictive diversity, aid choices need to drive improvements in individual accuracy by overwhelmingly favoring the most accurate aid.

Table \ref{tab:TableCPI} illustrates the results in greater detail. For the November 2021 collection, metawisdom emerged due to improvements in the mean prediction for each aid, suggesting there might have been some level of metacognitive matching whereby participants identified aids that would be more useful to them personally. This pattern does not extend to the later collections, which undermines such a claim. However, there was a fairly consistent distribution in aid choice. Despite being the most accurate aid, the \emph{Fed Statement} was chosen at about the same rate as \emph{Components}, the least accurate aid. Participants understandably preferred the \emph{Predictive Model} across all three collections because of its ease of application, but predictions made using the the Predictive Model varied across all data collections; for November 2021, the crowd correctly predicted the direction of change (an increase, although the magnitude was too large), but the other data collection samples incorrectly predicted the direction of change (increase, then decrease, while the actual changes were decrease and then increase, respectively). 

\begin{table}[h!]
\caption{CPI Nowcasting Experiment: Detailed illustration of results. $N$ is the number of participants in the respective treatment. The reported CPI values appear under the relevant month in the leftmost column.}
\label{tab:TableCPI}
\centering
\small
\renewcommand{\arraystretch}{1.2}
\begin{tabular}[]{ @{\extracolsep{.1em} } c | c || c | c | c | c | c | c | c }
\shortstack{CPI Forecast\\Date and Value} & \shortstack{Choice\\Treatment} & \shortstack{Decision Aid\\Treatment} & N & \shortstack{Mean\\Estimate} & GSE & MSE \\
\hline
\hline
\multirow{9}{*}{\shortstack{November 2021\\$\text{CPI} = 6.8$}}
& \multirow{4}{*}{Assigned} & Predictive Model & 91 & 8.11 & 1.73 & 15.67 \\
& & Fed Statement & 91 & 6.61 & 0.04 & 1.48 \\
& & Components & 91 & 11.25 & 19.83 & 88.41 \\
\cline{3-7}
& & \textit{All} & 273 & 8.66 & 7.20 & 34.90 \\
\cline{2-7}
& \multirow{4}{*}{Choice} & Predictive Model & 163 & 7.54 & 0.55 & 5.01 \\
& & Fed Statement & 51 & 6.51 & 0.09 & 0.69 \\
& & Components & 76 & 9.88 & 9.49 & 74.75 \\
\cline{3-7}
& & \textit{All} & 290 & 7.97 & 1.38 & 22.53 \\
\cline{2-7}
& No Aid & - & 90 & 7.13 & 0.11 & 124.95 \\
\hline
\hline
\multirow{9}{*}{\shortstack{May 2023\\$\text{CPI} = 4.0$}}
& \multirow{4}{*}{Assigned} & Predictive Model & 92 & 5.53 & 2.34 & 42.02 \\
& & Fed Statement & 92 & 5.00 & 0.99 & 5.31 \\
& & Components & 92 & 6.09 & 4.39 & 18.24 \\
\cline{3-7}
& & \textit{All} & 276 & 5.54 & 2.57 & 21.85 \\
\cline{2-7}
& \multirow{4}{*}{Choice} & Predictive Model & 160 & 5.22 & 1.49 & 3.46 \\
& & Fed Statement & 68 & 5.05 & 1.1 & 2.47 \\
& & Components & 66 & 7.89 & 15.14 & 70.31 \\
\cline{3-7}
& & \textit{All} & 294 & 5.78 & 3.17 & 18.24 \\
\cline{2-7}
& No Aid & - & 95 & 5.81 & 3.27 & 35.53 \\
\hline
\hline
\multirow{9}{*}{\shortstack{August 2023\\$\text{CPI} = 3.7$}}
& \multirow{4}{*}{Assigned} & Predictive Model & 85 & 3.54 & 0.03 & 3.08 \\
& & Fed Statement & 85 & 3.68 & 0.00 & 0.69 \\
& & Components & 85 & 6.76 & 9.37 & 60.92 \\
\cline{3-7}
& & \textit{All} & 255 & 4.66 & 3.13 & 21.56 \\
\cline{2-7}
& \multirow{4}{*}{Choice} & Predictive Model & 185 & 3.19 & 0.26 & 2.03 \\
& & Fed Statement & 51 & 3.92 & 0.05 & 1.72 \\
& & Components & 48 & 5.47 & 3.15 & 10.95 \\
\cline{3-7}
& & \textit{All} & 284 & 3.71 & 0.00 & 3.48 \\
\cline{2-7}
& No Aid & - & 99 & 5.64 & 3.74 & 67.52 \\
\hline
\end{tabular}
\end{table}
Our preregistered hypothesis stating the existence of metawisdom is supported for two of the three collection periods. However, it does not ultimately occur through a distribution of aid use that is inversely proportional to the aids' errors. Nonetheless, the consistent preference for the second most accurate aid combined with predictive diversity from using the other two aids twice led to a positive effect of choice.

\subsection{Study 3: Coin Count Estimation Task}
The design of the last study is broadly similar to the first two studies, but it has several additional elements designed to reveal more about the sources and robustness of metawisdom\footnote{The preregistration is available at \url{https://osf.io/bkhy7/?view_only=e57a85f51e70430c9a5a1bfef5c73360}}. Analogous to the first study, the task is a visual quantity estimation; in this case, the number of nickel coins in a glass cylinder\footnote{Because of its constant width, a cylinder is likely to be easier to reason about than the bowl in Study 1. However, nickels are likely a more challenging object than candies because their shape can lead to variably tight or loose packing. Taken together, we believe the difficulty of this prediction task is similar to Study 1}. However, before participants are aware of the presence of utilizing decision aids, we ask participants to make an unaided prediction so that we can assess the within-subject effect of aid use and whether aid choice is correlated with general predictive acumen. Additionally, the estimate is a distribution of possible values---a design choice discussed below. Finally, we have participants rank their preferences for aids according to a given probability of viewing them (70\%, 20\%, 10\%) in order to explore the effects of not receiving one's most preferred aid.

There are two primary reasons we elicit the estimation as a distribution of values \citep{morris_web-based_2014,falconer_methods_2022}. The first is to encourage more careful thinking throughout the process. The procedure first asks for participants' ``best guess,'' and then the lowest and highest ``number of coins you think might be possible.'' These questions should prompt additional reasoning that could potentially lead to more refined estimates. Using these values, we define a normal distribution in which the best guess is the mean value, and the extrema furthest from the mean are the values at three standard deviations away. Participants are then given an interactive visualization of that distribution, which they can further shape interactively (by clicking and dragging points) to their liking. Crucially, the interface visually links the distribution to the accuracy bonus, which is a function of the density of their distribution at the true value. The maximum bonus---a distribution of minimum variance centered on the true value---is \$50. By linking accuracy bonuses to a distribution, participants are incentivized to contemplate sources of uncertainty in their reasoning, perhaps leading to further refinement of the estimate. These data will also allow us to better understand the nature of the effects of individual aids. Study 3 consisted of 728 participants, and full details of the study design are provided in the Supplemental Materials Appendix \ref{sm: study3}. 

\subsection{Study 3 Results}
\begin{figure}
\centering
\includegraphics[scale=.3]{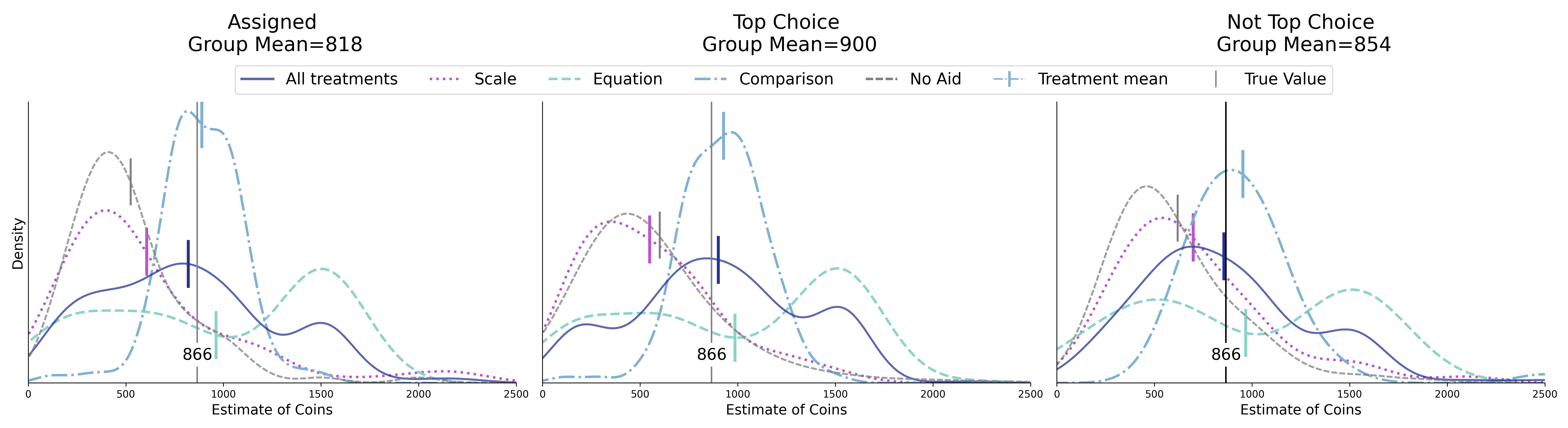} 
\caption{Study 3: Certainty in Estimation Tasks Experiment. The kernel density estimates (KDE) for the estimation distribution for the ``Assigned'' experimental treatment (left), ``Top Choice'' experimental treatment (middle), and ``Not Top Choice'' experimental treatment. Each panel contains KDEs for each of the three decision aids. The vertical line is the true value, 866, and the vertical hashes on the KDEs denote the crowd estimate for participants using the specific decision aid.}
\label{fig:coin_choice}
\end{figure}

Figure \ref{fig:coin_choice} uses KDEs to visualize the estimation distributions of different crowds. As in the previous studies, there is a panel for the Assigned condition. Here, however, we have separate panels for those in the Choice condition who viewed their most preferred aid (\textit{Top Choice}) and those who did not (\textit{Not Top Choice}). We use the mean of a participant's distribution as the contributed estimate to the crowd for all crowd-based analyses below. 

To assess the hypothesis that there is an information effect of this set of aids, we first confirm that the distributions of initial, unaided estimates for participants in the Assigned and Choice conditions are not different. The means of the initial estimates are 525.1 and 557.63, respectively, and we fail to reject the hypotheses that i) means are different (Welsh T-test, p=.242), and ii) the distributions are different (Kolmogorov-Smirnov test, p=.713). The average of the pooled initial estimates is 541.26, with a standard deviation of 339.01 and a skew of 2.17. The average standard deviation of the estimation distribution is 50.38\footnote{Two responses had very large standard deviation values and we drop them for this calculation, but not the analysis of means}, which implies participants supplied a range of almost 300 units between their lowest and highest ``possible'' values on average. 

We compare the crowd estimate of pooled initial estimates to the crowd estimate for the Assigned condition and, consistent with our preregistered hypothesis testing the existence of an information effect, find they are significantly different ($p<0.001$, Welsh's T). We also find evidence for metawisdom of the crowd also referred to as the choice effect. The crowd estimate in the Choice condition is 886, an error of 20, versus 818 for the Assigned condition, an error of 48. These means are significantly different ($p<0.05$, Welsh's T). When only those receiving their most preferred aid are considered, the crowd estimate increases to 900, which is still significantly different from the Assigned condition ($p<0.05$, Welsh's T).

Table \ref{tab:Table_coins} provides more details, including the number of participants using each aid. The Top Choice crowd favored the \emph{comparison} and \emph{equation} aids at similar rates (42\% compared to 43\% of group) in contrast to Study 1 where the \emph{equation} aid was selected at a greater rate (55.5\% compared to 33.5\%), suggesting that the additional elements of the design may have influenced the aid choices.  Interestingly, the Not Top Choice crowd is the most accurate because the comparatively heavy use of the \emph{scale} aid (40\% and majority selection for the group) provides substantial predictive diversity through error cancellation since it once again displayed an underestimation bias compared to other aids.

\begin{table}
\caption{Coin Count Estimation Task: Detailed breakdown of results. $N$ is the number of participants in the respective treatment.}
\label{tab:Table_coins}
\centering
\small
\renewcommand{\arraystretch}{1.2}
\begin{tabular}[]{ @{\extracolsep{.1em} } c | c || c | c | c | c }
Choice & Decision Aid & N & Mean & Group Squared Error & Mean Squared Error\\
Treatment & Treatment & & Estimate & (GSE) & (MSE) \\

\hline
\hline

\multirow{5}{*}
{Assigned } & {Scale} & 100 & 606 & 67636 & 272921\\
&{Equation} & 100 & 962 & 9130 & 309455  \\
&{Comparison} & 100 & 889 &537 & 41265\\
\cline{2-6}
&{\textit{All}} & 300 & 818 & 2220 & 207881\\
&{\textit{Initial}} & 300 & 525 & 116204 & 224858\\

\cline{1-6}

\multirow{5}{*}
{Top Choice} & {Scale} & 41 & 548 & 101124 & 211856\\
&{Equation} & 130 & 985 & 14161 & 328736\\
&{Comparison} & 127 & 927 & 3721 & 47781\\
\cline{2-6}
&{\textit{All}} & 298 & 900 & 1156 & 192919\\
&{\textit{Initial}} & 298 & 601 & 70225 & 514934\\
\cline{1-6}

& {Scale} & 52 & 698 & 28082 & 172981\\
\multirow{3}{2.5cm}{Not Top Choice} &{Equation} & 33 & 968 & 10342 & 301431\\
&{Comparison} & 45 & 953 & 7565 & 101971\\
\cline{2-6}
&{\textit{All}} & 130 & 854 & 121 & 181007\\
&{\textit{Initial}} & 130 & 619 & 61009 & 213973\\
\cline{1-6}

\hline
\end{tabular}
\end{table}
To better understand the origins of the metawisdom of the crowd effect found here, we test whether there is a relationship between the accuracy of the initial estimates and the choice of aids. It is possible that the judgment tasks are related (e.g., those who have a more accurate initial estimation model have better insight into which aids would actually improve their estimates) or correlated through a latent factor, such as thoughtfulness or general intelligence. We tested this using a multinomial logistic regression to predict the choice of aid as a function of the error size and standard deviation of the initial estimate. We fail to reject the null hypothesis that choice is independent of the prediction error and prediction standard deviation (S.D.). The results in Table \ref{tab:MNLogit} indicate that decision aid choice is not related to the accuracy of the initial prediction.
 
\begin{table}[h!]
\centering
\caption{Multinomial Logit of Aid Choice}
\label{tab:MNLogit}
\begin{tabular}{c|c|c}
\hline
& \textit{Equation} & \textit{Comparison}\\
\hline
Prediction Error & -0.00005& 0.0001\\
& (0.001)& (0.001) \\
Prediction S.D.    & -0.0003& 0.0011\\
& (0.002)& (0.002)  \\
Constant & 1.1771***& 1.0287***\\
& (0.278)& (0.264) \\
\hline
\multicolumn{3}{c}{Note: Reference level is Scale. Pseudo-$R^2$:0.004, N=428}\\ 
\multicolumn{3}{c}{Standard deviations are in parentheses. *** $p<0.001$}\\
\end{tabular}
\end{table}

Nonetheless, the utilized decision aid is predictive of the accuracy of the final estimate when controlling for the accuracy of the initial, unaided estimate. We test this separately for those who were assigned a decision aid and those who had the opportunity to choose an aid (\textit{Chose}=1). We further split the latter into those who viewed their top-ranked aid (\textit{Received Top Choice}=1) and those who did not. For the choice of aids, the \emph{scale} aid is again the reference category. The results of an ordinary least squares regression predicting individuals' absolute error appears in Table \ref{tab:OLS}.

\begin{table}[h!]
\centering
\caption{OLS, Predictors of Final Accuracy}
\label{tab:OLS}

\begin{tabular}{c|c}
\hline
Initial Prediction Error & 0.124***\\
& (0.019)\\
Initial Prediction S.D. & -0.015\\
& (0.017)\\
Viewed Equation Aid& 102.994***\\
& (19.916)\\
Viewed Comparison Aid & -249.270***\\
& (19.627)\\
Chose Aid & -20.5215 \\
& (21.474)\\
Received Top Choice & 14.428 \\
& (21.993)\\
Constant & 371.615***\\
& (18.514)\\
\hline
$N$ & 728\\
$R^2$ & 0.384\\
\hline
\multicolumn{2}{c}{Standard deviations are in parentheses. * $p<.1 $; **$p<.05$; *** $p<0.01$}\\
\end{tabular}
\end{table}
While the error on the initial, unaided estimate is predictive of accuracy on the final prediction, the effect size is minimal when compared to the effect of the aid that individuals actually viewed. Seeing the \textit{comparison} aid has a large negative effect on the size of final prediction error relative to the \textit{scale} aid. Seeing the \textit{equation} aid substantially increases the size of error; this aid has large, but diverse errors, stemming from the fact that participants may have used the wrong formula (e.g., they fail to add a packing coefficient). 

Having the opportunity to choose any aid, or to view one's top-ranked one, does not significantly alter individual errors. When combined with the aid-choice result, this emphasizes the curious nature of metawisdom. The individual choice of aid does not drive accuracy, which is readily observable in Table \ref{tab:Table_coins} because the mean squared errors increase for both the equation and comparison aids. Yet at the crowd level, the pattern of choices introduces vital predictive diversity that reduces the crowd error.  

\section{Discussion and Conclusion}
We found evidence for a metawisdom of the crowd effect in three different experimental tasks. The information present in the aids improved group estimates when participants were assigned to decision aids, but giving participants the choice of aids further improved group accuracy in most instances. These results show that the choice of decision aids can be a crucial decision that determines the accuracy of crowds on the primary decision task. In prior research, the primary estimation task has traditionally been the only decision under investigation. Because decision support systems are common, we argue that the decision of which aids to attend to is an important aspect of group decision-making. It has the potential to harm group decisions by diminishing the predictive diversity that underpins the wisdom of the crowd phenomenon. However, our studies provide evidence that the choice of decision aids can in fact contribute vital predictive diversity.

In designing the decision aids for our experiments, we sought to appeal to different cognitive styles or frames in an attempt to understand whether such aids systematically alter estimates. However, understanding the metacognitive processes deployed during aid choice and how individuals then reasoned with aids is beyond the scope of this paper. This limitation is in part because the outcomes of the metacognitive processes will depend on the collection of aids available for a given task, which will vary depending on contextual features. However, we now discuss several general ways in which the collection of aids might influence the metawisdom of the crowd.

First, we note that our theoretical model of metawisdom considers that the optimal distribution of aids will be dependent upon the biases and variances of the aids themselves. There are different cases where aid biases and variances may have different relationships with one another, and metawisdom may not be able to emerge under certain conditions. Specifically, metawisdom is a function of aid characteristics, and a collection of aids may not contribute additional variance beyond the variance of an individual estimator. In such cases, the information effect of the collection is approximately equal to the error of an unaided group of estimators (control in our empirical studies), and no further improvement is possible through aid use. If the choice of aid \emph{does not reduce} crowd accuracy, it suggests a type of collective wisdom regarding the choice of aid, but would not meet our definition of metawisdom since a crowd accuracy improvement is not observed. This is an artifact of how we define the default choice treatment but absent a strong justification for an alternative default treatment, random assignment appears to be a conservative and natural baseline for comparison.

The potential of a crowd with equal assignment to aids having zero group error raises the related question of whether the metawisdom we observed depends on the presence of aids that harm accuracy and whose presence contributed to the error in the crowd with Assignment to aids. In all experiments, participants chose the aid with the lowest group accuracy at a lower frequency than random (\emph{scale} aid in Study 1 and Study 3, and \emph{components} aid in Study 2), and this was an important source of improvement in group accuracy. Accordingly, a possible explanation of the observed metawisdom is that many participants avoided the aid that contributed most to the error in the Assigned condition. Is the availability of aids that harm accuracy necessary for a group to exhibit metawisdom? While the logic of avoiding inaccurate aids would be a powerful and intuitive example of metawisdom, we observed a more complex dynamic. In the experiments, the crowd members often failed to use the \emph{most} accurate aids as well. In fact, in Study 2, fewer individuals chose to view the most accurate aid than did the most inaccurate. So while avoiding the most inaccurate aid will decrease individual error and predictive diversity, we also observed an increase in individual errors and predictive diversity originating from other aspects of the collective choice of aids.

This dynamic highlights the fact that understanding the origins of metawisdom requires knowledge about the full collection of aids because little can be said about the relationship between individual aids and group accuracy without the aid context and observation of actual choices. For instance, the crowd completely avoiding the least accurate aid could reduce overall accuracy by removing predictive diversity. This can be seen in Study 1; if all participants had chosen the \emph{equation} and \emph{comparison} aids instead of the most inaccurate \emph{scale} aid, and the effects on errors of the former two aids under choice stayed the same, the overall crowd accuracy would have declined. Thus, the fact that some participants used the \emph{scale} aid increased crowd accuracy by adding predictive diversity more than it harmed individual accuracy. This pattern was not observed in the CPI Nowcasting experiment (Study 2), and it would have been impossible without choice simultaneously causing participants using one of the aids to systematically underestimate the true value. On the other hand, had choice not been associated with improvements in group accuracy for the Choice crowds, the observed selection tendency away from both the most and least accurate aids would have resulted in a group estimate of 8.68 for the Choice condition, slightly worse than the 8.66 of the Assigned treatment\footnote{$\frac{51}{290}(6.61) + \frac{163}{290}(8.13) + \frac{76}{290}(11.25) = 8.68$}.

These notably different consequences of individuals' choice of decision aids demonstrate the complex relationships between the set of aids, choice, and crowd accuracy. Both the set of available aids and the specific choices simultaneously affect crowd accuracy, and little can be said about the former without observing the latter. In scenarios where a controlling entity has authority over the provisioning of decision aids made available to decision-makers, one might hope that testing aids for accuracy and then removing the least accurate ones would improve crowd accuracy, but the dynamics we observe show that need not be the case. Even the use of behavioral nudges could produce unintended consequences by reducing the use of aids that enhance the predictive diversity of the crowd.

These types of scenarios abound. Some are common tasks related to making hiring or admission decisions through committee deliberations. Others are broader and less frequent, such as predicting economic trends or participating in democratic processes and elections. Individuals' choice from among the readily available aids is likely a foundational element in collective intelligence therein. This is not to diminish the individual consequences of relying on decision aids of poor quality, but the phenomenon of metawisdom suggests that decision aids of variable quality can still be associated with high-quality group decision-making in aggregate.

Of course, the formation of judgments and decision-making does not end with the viewing of decision aids, as many people go on to learn the tentative judgments of others and update their own via this social information and influence. Decision-makers might also search for different decision aids after learning about the judgments of others. These indirect and direct social influence processes are intertwined, but there is potential for metawisdom to be an essential underlying mechanism for generating the overall error distribution, and thus will condition the role of direct social influence on final group accuracy.

In conclusion, our experiments provide evidence that individual choices regarding decision aids contribute to the accuracy of a group performing an estimation task. This accuracy did not occur through the straightforward mechanism of the crowd gravitating toward the most helpful (i.e. highest accuracy inducing) aids, but rather through the same basic pattern of decision accuracy that supports crowd wisdom more generally; some crowd members ``accurately'' identified the most helpful aid, but the bulk of the crowd erred by choosing the second most helpful aid. A remainder of the crowd chose poorly by using the most inaccurate aid. This dispersion of errors on the first choice mirrors that of crowds more generally, but incorporating valuable information and interpretative frames via decision aids also boosts individual accuracy on the primary decision.

The existence of multiple patterns of decision-aid choice that can enhance crowd accuracy has implications for how we approach the study of crowd wisdom in domains with decision-support architectures, including within organizations, markets, and democratic processes. Our results cannot directly address outcomes in those diverse settings due to a lack of empirical data on the composition of decision aids and actual choice behavior, but nonetheless, we provide a framework for analyses and a standard for comparison and design aspiration---metawisdom of the crowd.

\bibliographystyle{informs2014}
\bibliography{references}

\newpage

\renewcommand{\thesection}{\Alph{section}}
\renewcommand{\thesubsection}{A\arabic{subsection}}
\setcounter{figure}{0}
\renewcommand\thefigure{A\arabic{figure}}
\setcounter{table}{0}
\setcounter{equation}{0}
\renewcommand\thetable{A\arabic{table}}
\section*{Supplemental Materials}
\vspace{1em}
\subsection{Calculation of Optimal Aid Distribution} \label{sm: calculate}
Solving for the optimal distribution of aids requires formalizing the group squared error equation in terms of aid bias and variance as presented in the main text and below.Here in the appendix, we use the subscript $k$ to distinguish between a generic set of aids $k$ and the aid of interest $i$.

\begin{align}
\text{GSE}(C) &= \left(\sum_{k=1}^{n} p_k B_k\right)^2 + \sum_{k=1}^{n} p_k^2 \sigma_k^2
\end{align}

As a reminder, here are the key elements of the equation:

\begin{itemize}
\item $B_k$ is the bias of aid $k$
\item $\sum_{k=1}^{n} p_k B_k$ is the pooled bias 
\item $\sigma_k^2$ is the variance of aid $k$
\item $p_k$ is the proportion of a group using the aid $k$ and has the following constraint: $\sum_{k=1}^{n} p_k = 1$
\end{itemize}

Due to the constraint on the sum of aid proportions, the problem can be formulated as a constrained optimization problem whose solution may be developed using the Lagrange Multipliers methodology. Lagrange Multipliers require calculating partial derivatives to optimize a function under constraints \citepsupp{bertsekas2014constrained,everett1963generalized}; in our case, we solve for the optimal proportion of a given aid, $p_i$. The Lagrangian equation is as follows: 
\begin{align}
L(p_1, p_2, \ldots, p_n, \lambda) = \left(\sum_{k=1}^{n} p_k B_k\right)^2 + \sum_{k=1}^{n} p_k^2 \sigma_k^2 - \lambda \left(\sum_{k=1}^{n} p_k - 1 \right)
\end{align}

In the calculation of the partial derivatives with respect to $p_i$ ($i= 1, 2, ..., n$), each of the expressions for pooled bias, variance, and the proportional constraint must be differentiated. We use notation to differentiate for a specific aid, $p_i$, instead of the set of aids, $p_k$, because the model must determine the optimal proportion for each aid, and the solution demonstrates a general optimization.

\begin{align}
\frac{\partial L}{\partial p_i} &= \frac{\partial}{\partial p_i}\left[\left(\sum_{k=1}^{n} p_k B_k\right)^2\right] + \frac{\partial}{\partial p_i}\left(\sum_{k=1}^{n} p_k^2 \sigma_k^2\right) - \lambda \frac{\partial}{\partial p_i} \left[\left(\sum_{k=1}^{n} p_k - 1 \right)\right]
\end{align}

\begin{align}
\frac{\partial L}{\partial p_i} &= 2\left(\sum_{k=1}^{n} p_k B_k\right)\cdot B_i + 2p_i \sigma_i^2 - \lambda
\end{align}

We then set the previous expression to zero and solve for $p_i$:

\begin{align}
2\left(\sum_{k=1}^{n} p_k B_k\right)\cdot B_i + 2p_i \sigma_i^2 - \lambda &= 0
\end{align}

\begin{align}
p_i &= \frac{\lambda - 2\left(\sum_{k=1}^{n} p_k B_k\right)\cdot B_i} {2 \sigma_i^2}
\end{align}

Since the pooled bias term includes a reference to our focal aid proportion ($p_k$ contains $p_i$), that reference must be addressed to accurately express the optimal aid proportion:

\begin{align}
\sum_{k=1}^{n} p_k B_k &= \sum_{k=1}^{n}\frac{\lambda - 2\left(\sum_{k=1}^{n} p_k B_k\right)\cdot B_k} {2\sigma_k^2} \cdot B_k =
\frac{\lambda}{2} \left(\sum_{k=1}^{n} \frac{B_k}{\sigma_k^2}\right) - \left(\sum_{k=1}^{n} p_k B_k\right)\left(\sum_{k=1}^{n} \frac{B_k^2}{\sigma_k^2}\right)
\end{align}

Isolating all terms in the expression that reference $p_k$:

\begin{align}
\left(\sum_{k=1}^{n} p_k B_k\right)\left( 1 + \sum_{k=1}^{n}\frac{B_k^2}{\sigma_k^2}\right) &= 
\frac{\lambda}{2} \left(\sum_{k=1}^{n} \frac{B_k}{\sigma_k^2}\right) 
\end{align}

Now, the pooled bias term is expressed without referencing aid proportions within the expression:

\begin{align}
\sum_{k=1}^{n} p_k B_k &= 
\frac{\lambda \left(\sum_{k=1}^{n} \frac{B_k}{\sigma_k^2}\right)}{2 \left( 1 + \sum_{k=1}^{n}\frac{B_k^2}{\sigma_k^2}\right)} 
\end{align}

To solve for $p_i$, we also have to solve for $\lambda$ using the defined constraint of $\sum_{k=1}^{n} p_k = 1$. Substituting for $p_k$ produces the following expression:

\begin{align}
\sum_{k=1}^{n} \frac{\lambda - 2\left(\sum_{k=1}^{n} p_k B_k\right)\cdot B_k} {2 \sigma_k^2} = 1
\end{align}

Then, without referencing aid proportions:
\begin{align}
\frac{\lambda}{2}\sum_{k=1}^{n} \frac{1}{\sigma_k^2} - \frac{\lambda \left(\sum_{k=1}^{n} \frac{B_k}{\sigma_k^2}\right)}{2 \left( 1 + \sum_{k=1}^{n}\frac{B_k^2}{\sigma_k^2}\right)} \sum_{k=1}^{n}\frac{B_k}{\sigma_k^2} = 1
\end{align}

Further manipulation:
\begin{align}
\lambda\left(\sum_{k=1}^{n} \frac{1}{\sigma_k^2} - \frac{\left(\sum_{k=1}^{n} \frac{B_k}{\sigma_k^2}\right)^2}{\left( 1 + \sum_{k=1}^{n}\frac{B_k^2}{\sigma_k^2}\right)}\right) = 2
\end{align}

Finally, we have an expression for $\lambda$:
\begin{align}
\lambda &= \frac{2}{\left(\sum_{k=1}^{n} \frac{1}{\sigma_k^2} - \frac{\left(\sum_{k=1}^{n} \frac{B_k}{\sigma_k^2}\right)^2}{\left( 1 + \sum_{k=1}^{n}\frac{B_k^2}{\sigma_k^2}\right)}\right)} = \frac{2\left(1 + \sum_{k=1}^{n}\frac{B_k^2}{\sigma_k^2} \right)}{\sum_{k=1}^{n} \frac{1}{\sigma_k^2} \left(1 + \sum_{k=1}^{n}\frac{B_k^2}{\sigma_k^2} \right) - \left(\sum_{k=1}^{n} \frac{B_k}{\sigma_k^2}\right)^2}
\end{align}

We can now present the final solution for $p_i$ and begin substituting other proportions and $\lambda$.

\begin{align}
p_i &= \frac{\lambda - 2\left(\sum_{k=1}^{n} p_k B_k\right)\cdot B_i} {2 \sigma_i^2}
\end{align}

The following expression emerges after substituting for the pooled biases and simplifying.
\begin{align}
p_i &= \frac{\lambda}{2\sigma_i^2}-\frac{B_i}{\sigma_i^2} \cdot
\frac{\lambda \left(\sum_{k=1}^{n} \frac{B_k}{\sigma_k^2}\right)}{2 \left( 1 + \sum_{k=1}^{n}\frac{B_k^2}{\sigma_k^2}\right)} = \frac{\lambda}{2\sigma_i^2} \left(1 - B_i \frac{\left(\sum_{k=1}^{n} \frac{B_k}{\sigma_k^2}\right)}{ \left( 1 + \sum_{k=1}^{n}\frac{B_k^2}{\sigma_k^2}\right)}\right)
\end{align}

Then, after substituting for $\lambda$, the expression has the following form.
\begin{align}
p_i &= \frac{2\left(1 + \sum_{k=1}^{n}\frac{B_k^2}{\sigma_k^2} \right)}{2\sigma_i^2 \left(\sum_{k=1}^{n} \frac{1}{\sigma_k^2}\right) \left(1 + \sum_{k=1}^{n}\frac{B_k^2}{\sigma_k^2} \right) - \left(\sum_{k=1}^{n} \frac{B_k}{\sigma_k^2}\right)^2} \left(1 - B_i \frac{\left(\sum_{k=1}^{n} \frac{B_k}{\sigma_k^2}\right)}{ \left( 1 + \sum_{k=1}^{n}\frac{B_k^2}{\sigma_k^2}\right)}\right) 
\end{align}

Lastly, the final form of the expression describing the optimal proportion of a decision aid.
\begin{align}
    p_i = \frac{1}{\sigma_i^2} \cdot \frac{\left(1 + \sum_{k=1}^{n} \frac{B_k^2}{\sigma_k^2}\right) - B_i \sum_{k=1}^{n} \frac{B_k}{\sigma_k^2}}{\left(\sum_{k=1}^{n} \frac{1}{\sigma_k^2}\right) \left(1 + \sum_{k=1}^{n} \frac{B_k^2}{\sigma_k^2}\right) - \left(\sum_{k=1}^{n} \frac{B_k}{\sigma_k^2}\right)^2}
\end{align}

\bibliographystylesupp{informs2014}
\bibliographysupp{supplement-2025_refs}

\subsection{Computational Model of Aid Use} \label{sm: computation}
To complement the analytical model of optimal aid distribution, we also developed a computational model that explores the relationship between distributions of decision-aid use and group accuracy. It is a minimal model that highlights our main intuition that if aid use is roughly inversely proportional to the mean prediction error induced by the aid, then choice-enabled crowds will be more accurate than crowds assigned randomly to aids. 

We conceive of decision aids as the output of a process that transforms a set of complex inputs to a more readily usable signal, often at the cost of not fully capturing the information available. This implies that using decision aids can lead to a consistent predictive bias for those who use them. For the model, we assume decision-makers have no private information and no interpretive variability; the signal given by the decision aid always results in the same prediction for every decision-maker using it. This allows us to define $x_i$ as the prediction associated with the $i^{\text{th}}$ decision-aid of $\textbf{M}$ such aids. We again denote the true value as $Y$. 

When $p_i$ is the proportion of the population using the $i^{\text{th}}$ decision aid, we can define a crowd estimate as:
\begin{align*}
    C = \sum_i^{\textbf{M}} p_ix_i
\end{align*}
Therefore a crowd's accuracy is exclusively a function of the relationship between $x_i$, the estimate of the $i^{\text{th}}$ decision aid, and $p_i$, the proportion of the crowd using that decision aid. 

The intuition behind existence of metawisdom of the crowd is that individuals should want to choose the aid that induces the most accurate estimate and their accuracy on this decision task might be distributed roughly in proportion to the accuracy of the aid. That is, the probability of choosing an aid is inversely proportional to the size of the squared error\footnote{Individuals likely think in terms of a signed, unsquared error, but our analysis revealed no meaningful performance difference from the squared error approach we take and a version with absolute errors.}, or $p_i \propto \frac{1}{(Y-x_i)^2}$. To model this, we define $z_{i}$ as the inverse of the proportion of the $i^\text{th}$ decision-aid's squared error and the sum of all decision aids' squared errors. 
\begin{align*}
     z_{i} = \frac{\sum^\textbf{M}_j(Y-x_j)^2}{(Y-x_i)^2}
\end{align*}
To obtain a weighting that is inversely proportional to the error, we normalize $z_i$:
\begin{align*}
    p^{\prime}_{i} = \frac{z_{i}}{\sum_j^{\textbf{M}}{z_{j}}},
\end{align*}
When aid use is allocated according to $p^\prime_i$, which we refer to as the $p^\prime$ allocation, the corresponding crowd estimate is:
\begin{align*}
    C^\prime = \sum_{i=1}^{\textbf{M}} p^\prime_i x_i.
\end{align*}
According to the Diversity Prediction Theorem \citepmod{hong_groups_2004}, there are two analytically distinct sources of crowd accuracy. One is the reliance on the most accurate aids, which reduces the average individual squared error. The other is the balancing out of errors through the incorporation of predictive diversity. The more distant an $x_i$ is from the group average, the larger the predictive diversity term, and because that term is subtracted from the other, the group error goes down. Thus, the average individual error can be quite large while the group is nonetheless completely accurate. This happens whenever the individual estimates suitably ``bracket'' the true value  \citepmod{davis-stober_when_2014}. 

The $p^{\prime}$ allocation prioritizes minimizing individual errors by favoring the most accurate aids, but introduces predictive diversity by allowing some weight on the other aids. To demonstrate the value of the predictive diversity created by the $p^\prime$ allocation, we interpolate between it and two baseline allocations. The first, $p^\text{equal}$, is simply equal proportions for each aid, or 
\begin{align*}
    p_i^\text{equal}= \frac{1}{\textbf{M}}.
\end{align*}
Alternatively, if the mean prediction of each aid is known, one could allocate all weight to the most accurate or \textit{best} aid, $p^\text{best}$:
\begin{align*}
    p_i^\text{best} = \begin{cases}
     1, &         i = \text{arg min }(Y-x_i)^2\\
            0, &         \text{otherwise}
    \end{cases}
\end{align*}
The crowd estimates associated with these two schemes are:\begin{align*}
C^\text{equal} = \sum_i^{\textbf{M}} p_i^\text{equal}\: x_i \text{, }
\end{align*}\begin{center}
    and
\end{center}\begin{align*}
C^\text{best} = \sum_i^{\textbf{M}} p_i^\text{best}\: x_i
\end{align*}
We interpolate between all three allocations by adding or subtracting a constant increment to all values in the $p^\prime$ vector of weights and normalizing it. Progressively adding an increment to each proportion shifts the allocation toward $p^\text{equal}$. Subtracting the constant eventually results in the aid with the smallest error being the only one in use (i.e., $p^\text{best}$). As an increment, we use $\gamma/5$, where $\gamma$ is equal to the \textit{second} largest proportion in $p^\prime$, because all weight on the second largest proportion needs to be removed in order for the \textit{best} aid to have the only positive weight. We number the incrementing steps using $\mathbf{x}$.
\begin{align*}
p^\star_{i,\mathbf{x}} = \mathbf{max}(p^\prime_i + \frac{\gamma}{5} \mathbf{x}, 0),
\end{align*}
\begin{align*}
p_{i,\mathbf{x}} = p^\star_{i,\mathbf{x}}/\sum_j^M p_{j,\mathbf{x}}^\star
\end{align*}
\begin{center}
    and
\end{center}
\begin{align*}
C^\mathbf{x} = \sum_{i=1}^\mathbf{M} p_{i,\mathbf{x}}\:x_i
\end{align*}
For example, let $p^\prime = [.55,.35,.10]$. Then $\gamma = .35$ and when $x=-1$, $p^\star_{i,-1}= [(.55-.07),(.35-.07), (.10-.07)]$ and $p_{i,-1} = [.608,.354,.038]$.  When $x=-2$, the least accurate aid has a weight of zero (i.e., $p_{3,-2}=0)$. Once $x=-5$, the most accurate aid has a weight of 1 (i.e., $p_{1,-5}=1$) and the allocation is equivalent to $p^\text{best}$. In this example,  the weights are approximately equal to a third at two decimal places (equivalent to $p^\text{equal}$) once $x\geq23$.

\subsubsection{Model Analysis }
The expected accuracy of crowd estimates depends on $\textbf{M}$, the number of aids, because increasing $\mathbf{M}$ increases i) the accuracy of the most accurate aid in expectation, and ii) the probability that there are aids with differently signed errors. The former implies a general increase in accuracy while the latter increases the potential to incorporate more diverse predictions. Therefore we conduct 1,000 trials for $\textbf{M} \in [2,5]$. In each trial, $\textbf{M}$ aid means (i.e., $x_i$) are sampled from $\mathcal{N}(Y=0,\sigma=1)$, meaning the true value is fixed at $0$. This set of aids is used to calculate $C^\mathbf{x}$ for all values of $\mathbf{x}$.

Given a set of $\textbf{M}$ aid means, we calculate the difference between the group squared error of $C^\mathbf{x}$ and the $C^\text{equal}$ crowd. This facilitates comparison across trials because accuracy depends on the composition of the aid-means in the set. Additionally, as a unit for comparison, we use the magnitude of the group squared error of the $C^\text{equal}$ crowd. Thus, the maximum increase in accuracy relative to the accuracy of $C^\text{equal}$ crowd is 1 unit because at that point the crowd estimate is equal to the true value, $Y$. In principle, decreases in accuracy can be much larger, but in practice are bounded to approximately -1 units.

\begin{figure}[h]
    \centering
    \includegraphics[width=\textwidth]{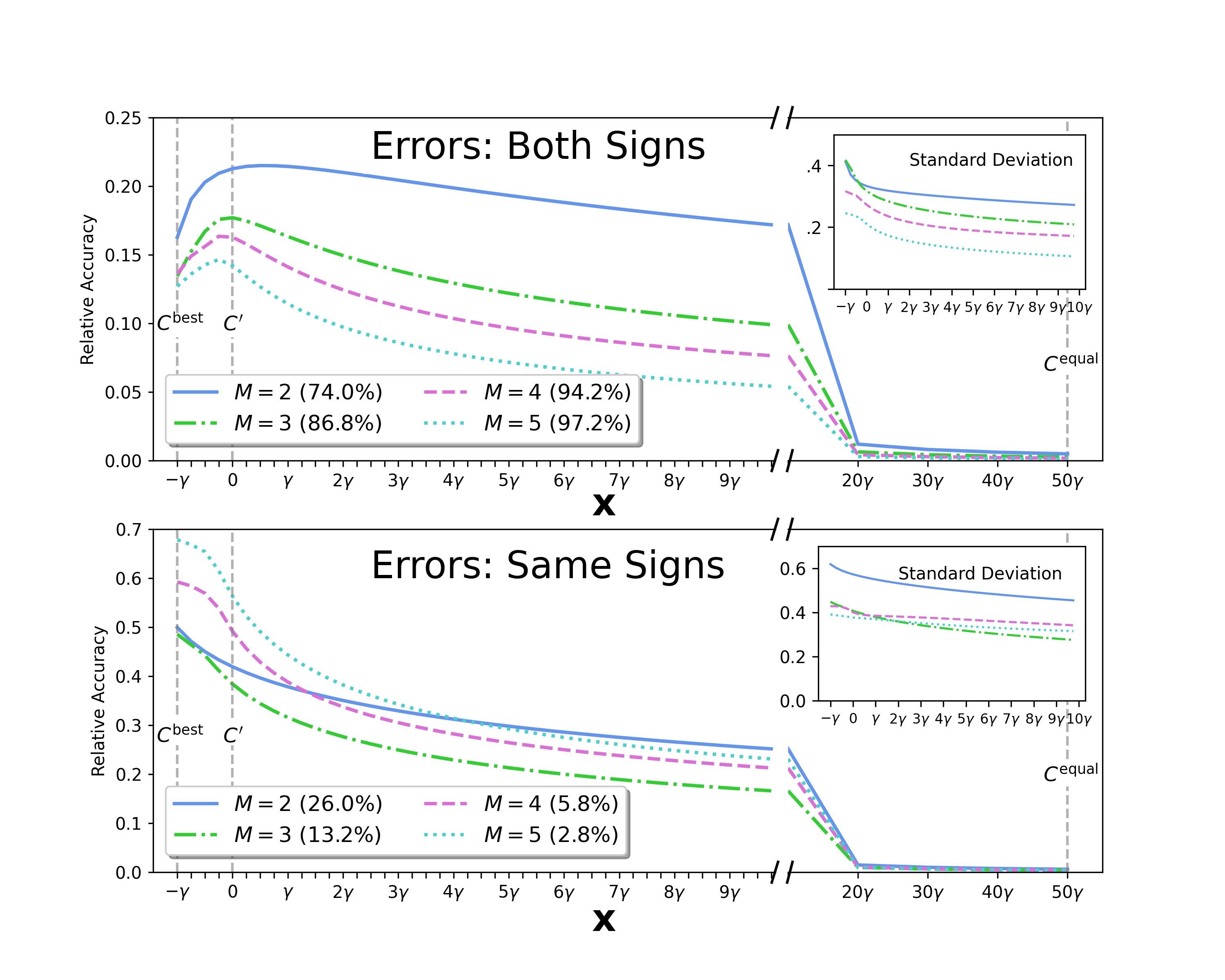}
    \caption{Minimal Model: Relative accuracy of $C^\mathbf{x}$ for a range of $\mathbf{x}$ that encompasses both $C^\text{best}$ and $C^\text{equal}$, both of which are noted as vertical dashed lines along with $C^\prime$. The top panel contains the trials for which the set of aids has errors of each sign. The bottom panel is for sets with the all the same sign. The legends report the percentage of all 1000 trials in the respective panel for each $\mathbf{M}$. The insets plot the standard deviations for the same trials. Note the discontinuity of scale for $\mathbf{x}$. }
    \label{fig:core-results}
\end{figure}

Figure \ref{fig:core-results} separates the accuracy results into panels by whether the randomly sampled aids have the same (positive or negative) signs on their errors (Top Panel: Both Signs, Bottom Panel: Same Signs). The observed percentages for the 1,000 trials per value of $\mathbf{M}$ are reported in the respective legends\footnote{When treated as a Bernoulli trial with $\mathbf{M}$ draws from the unit normal distribution, the probability of all draws having the same sign is just $(\frac{1}{2})^\mathbf{M}$.}. On the x-axis, we sweep $\mathbf{x}$ from -5 to 250 (i.e., $\gamma \in [-1,50]$, because each $\mathbf{x}$ increment is $\gamma/5$ units); note the change of scale for values greater than 50. The values at the far left ($\mathbf{x} = -5$) is the accuracy measure for $C^\text{best}$; the far right, for $C^\text{equal}$. The vertical dashed line at $\mathbf{x}=0$ corresponds to the group accuracy of the $C^\prime$ crowd. The standard deviations are visualized in the inset plots.

The $C^\prime$ crowd outperforms both $C^\text{equal}$ and $C^\text{best}$ in expectation when the set of aids have errors with both signs. The outperformance of $C^\text{best}$ by $C^\prime$ demonstrates the substantial predictive diversity the $p^\prime$ allocation brings to the crowd. In fact, for this method of interpolation, the $p^\prime$ allocation is nearly optimal. The peak relative performance of $C^\mathbf{x}$ is at approximately $C^\prime$. Even moving more of the allocation toward the less accurate aids can still allow the new crowd to outperform $C^\text{best}$. The standard deviations are sizable, however, as there exist sets of aids for which $C^\text{equal}$ is highly accurate (e.g., $x_i = (-0.4,0.15,0.25)$, meaning $C^\text{equal} = Y$) and favoring any aid can only make the group error worse.

When all aids have the same sign, $C^\text{best}$ is the top-performing allocation. This is necessarily the case because allocating use toward any other aid increases the mean squared error faster than it can increase the predictive diversity. However, in expectation, the difference between the relative accuracy improvement of $C^\text{best}$ and $C^\prime$ is substantially less than the difference between $C^\text{equal}$ and $C^\prime$. 

These results indicate that an allocation for the use of decision aids that is inversely proportional to the squared error of the aids (i.e., $p^\prime$) would generally be superior to an \textit{equal} allocation (i.e., $p^\text{equal})$, in spite of the latter often containing a substantial predictive diversity itself. Furthermore, inverse proportionality can often outperform distributing everyone to the single most accurate aid, $p^\text{best}$. In fact, given the relatively low frequency with which all aids have the same sign, $p^\prime$ performs as well as or better than $p^\text{best}$ in expectation for all randomly created aid sets.

\begin{figure}[h]
    \centering
    \includegraphics[scale=.5]{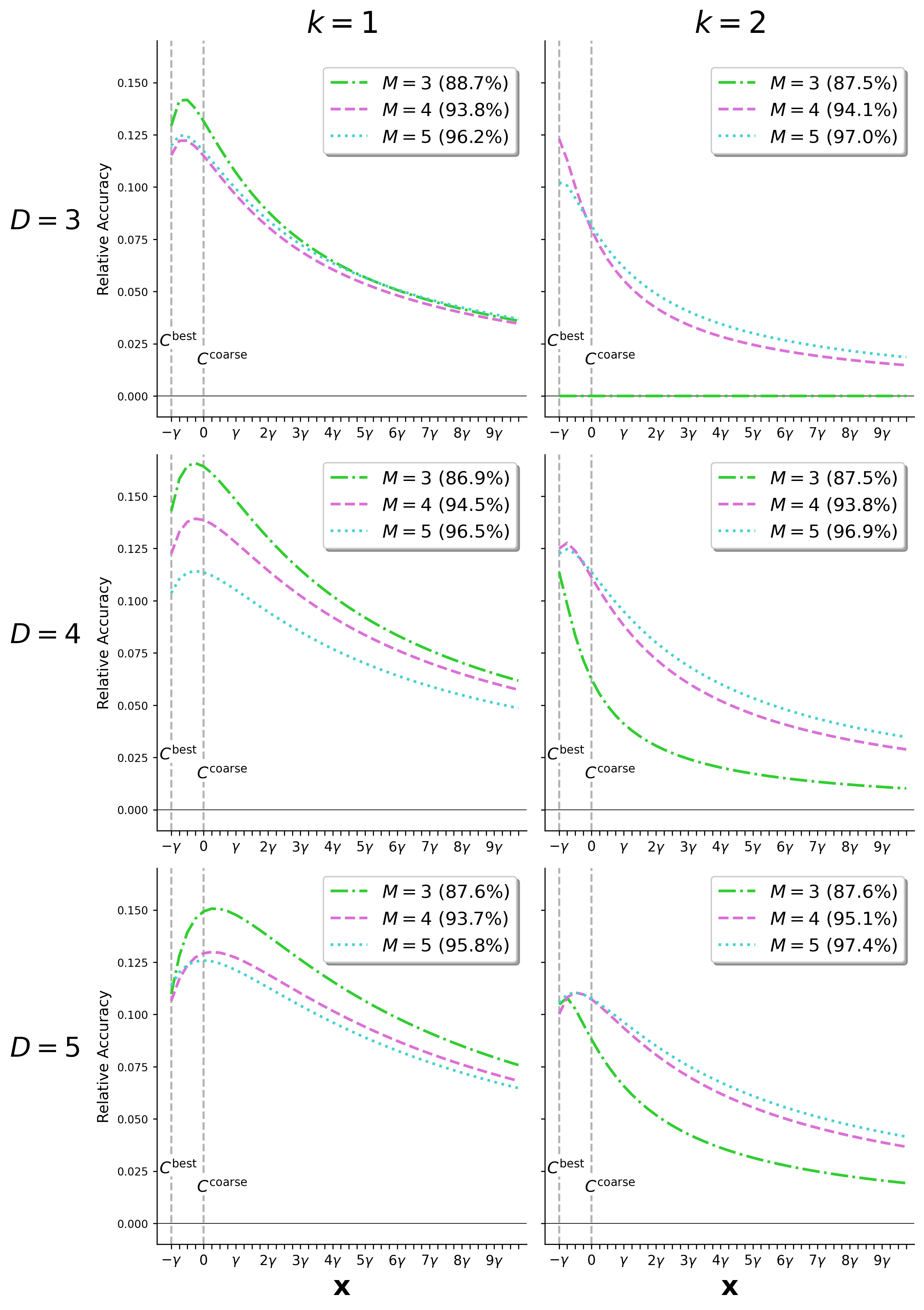} 
    \caption{Coarse allocation model: Accuracy of $C^\mathbf{x}$ for $k \in [1,2]$ and $D \in [3,5]$. Only the trials with a set of aids that has at least one aid with an error of each sign are reported; the percentage of all 1000 trials appear in the legends.  $C^\text{best}$ and $C^\text{coarse}$ are noted as vertical dashed lines.}
    \label{fig:coarse_model}
\end{figure}

Interpolating as we do indicates the rule of inverse proportionality is robust to deviations away from the formulaic distribution, but our interpolation method covers only a small fraction of the space of possible distributions. To further demonstrate the robustness of the intuition behind the rule, we also analyze a set of rules that also favors more accurate aids, but with less precision. The formulas below represent a family of rules that divides the aids into a favored group and an unfavored group, where the favored group contains the top $k$ most accurate aids. By extension, the unfavored group contains $\mathbf{M}-k$ aids. Constant $D$ controls the fraction of the total usage each group gets; the higher $D$ is, the more weight ( $p^{\text{favored}}$) is given to aids in the favored group. Those in the unfavored group received $p^{\text{unfavored}}$. 
\begin{align*}
p^\text{favored} = \frac{D-1}{Dk},\\
\\
p^\text{unfavored} = \frac{1}{D(M-k)}
\end{align*}
We refer to the resulting allocation as $p^\text{coarse}$ and a crowd using it has prediction $C^\text{coarse}$. For example, for $D=5$ and $k=2$, a set of aids with means $x_i=[-0.4,\;-0.05,\; 0.1,\; 0.3]$ would result in $p^\text{coarse}= [\frac{1}{10}, \frac{4}{10}, \frac{4}{10}, \frac{1}{10}]$.  Figure \ref{fig:coarse_model} contains six panels for the combinations of $k\in[1,2]$ and $D\in[3,5]$, only for the trials with aids with errors of different signs. For $k=1$, varying $D$ has little effect on relative performance; . It performs at least as well as $C^\text{best}$ in expectation and substantially better than $C^\text{equal}$. When $k=2$, $C^\text{coarse}$ underperforms $C^\text{best}$ while still outperforming $C^\text{equal}$ in expectation. 

The $p^\text{coarse}$ allocation demonstrates that metawisdom is possible in expectation when inverse proportionality to the squared error is loosely define. There is a wide range of allocation patterns that would result in a crowd outperforming an equal allocation to aids. Many of those allocations can outperform using only the most accurate aid as well. The natural next question is how actual crowds tend to choose aids and whether those patterns exhibit metawisdom of the crowd. We now approach answering that question experimentally.

\bibliographystylemod{informs2014}
\bibliographymod{references}

\subsection{Study 1 details} \label{sm: study1}

The results discussed in the main text do not cover the full design of this exploratory study. First, there was a fourth choice treatment condition, \textit{Multiple choice}. Participants in that condition could view all aids as many times as they would like by navigating backward to the aid choice page. The outcomes for this condition are in Table \ref{tab:SI_multi}. The crowd estimate was more accurate and statistically different from the Assigned condition (H2: A choice effect, or metawisdom of the crowd). The estimate was not meaningfully different from the Choice condition so we omit Multiple Choice from the main text.  36 (9\%) participants viewed two aids and 114 (29\%) viewed all three.

\begin{table}[b]
\caption{Coin Count Estimation Task: Detailed breakdown of results for the Multiple Choice treatment. Aid use counts are determined by the last aid each participant used. GSE is the group squared error. MSE is the mean squared error.}
\label{tab:SI_multi}
\centering
\small
\renewcommand{\arraystretch}{1.2}\begin{tabular}[]{ c | c || c | c | c | c }
Crowd & Information\\
{Treatment} & {Treatment} & $N$ & Mean Estimate & \textit{GSE} & \textit{MSE} \\

\hline
& {\scriptsize Scale} & 41 & 318 & 29033 & 71911\\
{\small Multiple}&{\scriptsize Equation} & 234 & 578 & 8118 & 86802 \\
{\small Choice}&{\scriptsize Comparison} & 123 & 528 & 1625 & 23027\\
&{\scriptsize \textit{All}} & 398 & 536 & 2293 & 65559 \\
\hline
\end{tabular}
\end{table}

The exploratory study was in part motivated by the belief it is likely that decision aids frame decisions in ways that bias the interpretation of the information. To assess whether framing effects do indeed play a role, we deployed a ``within-crowd'' design that compares a distribution of estimates for nearly identical tasks. The analysis in the main text focused on estimating the count of hard candies, but all participants of Study 1 also estimated a count for a second type of candy, an animal shaped gummy, after the hard candies. This happened after participants did four filler tasks and after having been shown a randomized decision aid. We implemented the design for all three choice conditions (i.e., Assigned, Single Choice and Multiple Choice) but present only the results of the Assigned condition because the results are very similar across conditions. The combinations of the two tasks with three different treatments per task results in nine different treatment groups (e.g., scale on the first task and equation on the second task versus equation on the first task and scale on the second.)

We treat each combination as a crowd itself and look at the differences in the distribution of estimates after each exposure. The design is like a within-subject design, but we are not presently interested in individual level effects. Instead, we are interested in the change in the distribution of estimates within the crowd. Those differences would be evidence of priming or framing effects because participants can apply anything they learned in the first task to the second one. However, exposure to a different decision aid could cause individuals to see the problem differently and, presumably, estimate differently despite the true counts being only 5 candies different (488 and 493).
\begin{figure*}[h]
\centering
\includegraphics[scale=.4]{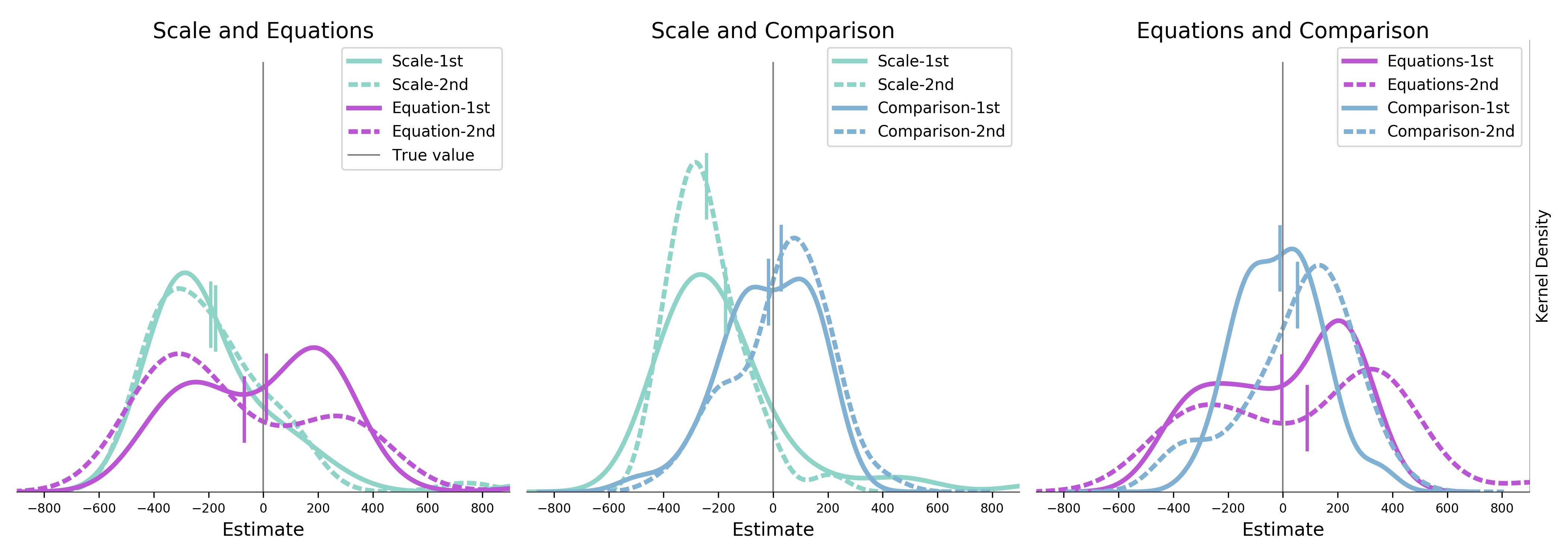} 
\caption{Aid Framing: Mean-centered kernel density estimates (KDE) for the two estimation tasks completed by participants in the Assigned condition (participants were randomly assigned information aids for both tasks). The solid line KDE corresponds to estimates when the information aid appeared in the first task and the dashed line for the second task. The colors correspond to the different aids: teal for scale, magenta for equation, and powder blue for comparison. Vertical hashes mark the mean of the corresponding distribution.}
\label{fig:gummy-mm-one-by-three}
\end{figure*}

As a baseline, we argue that if only the informational content of the aids matter, participants will produce the same distribution of estimates, regardless of the order in which the aids were seen. We do not observe this. Figure \ref{fig:gummy-mm-one-by-three} shows the mean-centered kernel density estimations (KDEs) of the distribution of estimates for the six subgroups that received exposures to two different aids. These subgroups are grouped into panels of the same two exposures, but in different orders. For example, the left panel contains the KDEs for the two treatment groups that viewed both the scale and equation aids but in different orders. The two distributions for the scale aid are nearly indistinguishable despite the fact that one group had used the equation aid previously. This suggests that participants are not transferring information from the use of the equation aid to the use of the scale aid later and the information and frames carried by the scale aid alone drive the estimates. (Participants did not learn the outcome of the first task until the following day to avoid the possibility of the true value being shared on online forums. Also such feedback could alter subsequent decisions by adding additional information, thereby undermining experimenter control over the treatment conditions.)  Contrast that with the distributions for the equation aid in that same panel; having first seen the scale shifts the estimates to the left relative to those who first used the equation aids.

Figure \ref{fig:gummy-mm-one-by-three} provides some visual evidence of ordering effects (i.e., the KDEs for the same decision aid can vary somewhat within and across panels), but also evidence that estimates made after exposure to a given aid in the second task are similar to those made after exposure in the first task, indicating that the aid in the second task supersedes the information or prime from the first task. Only one of the six subgroups tested significantly for the different in means for the same aids (Comparisons in the right panel, p=0.08, Welsh's T-test). Taken together, this analysis shows that participants' use of decision aids is quite complex. If estimates for the second task were similar, one could argue that decision aids influence decisions primarily via the information they carry and that once two participants has seen both aids, regardless of the ordering, they are at information parity. We do not observe that here and tentatively conclude that decision aids bias decisions at least in part through the framing of information at the expense of other interpretative frames.

\subsubsection{Study 1 materials}
The text for the second, gummies task was identical except for the substitution of the word ``gummies'' for ``candies.'' The bowls in the gummy images were the same.
\begin{figure}[H] 
\centering
\includegraphics[width=.5\linewidth]{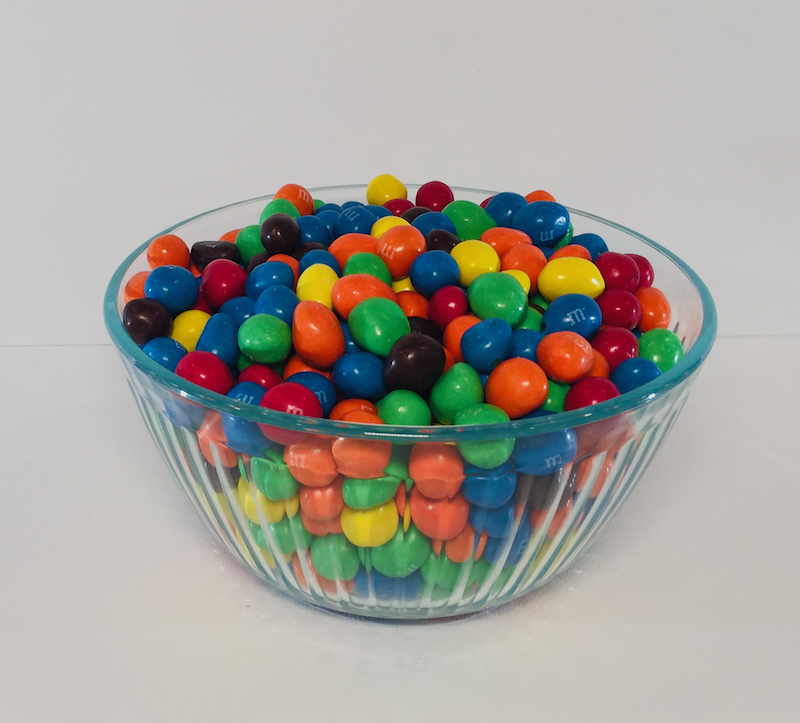}
\caption{The primary image for the estimation task. The associated text says ``For this task you need to estimate the number of candies in the bowl in the picture below. REMINDER: If your answer is within 5\% of the correct answer you'll get a \$2.00 bonus AND the best estimate will receive a \$25.00 bonus too. In the case of ties, the winners will split a \$50 reward pot.''} 
\label{fig:candies}
\end{figure}

The Equation aid displayed the same primary image as above and contained the following text:
Here is the same picture of the bowl as on the first page. To help you estimate, here are two volumes to consider.
\begin{itemize}    
    \item{The volume of the bowl to the brim is approximately 1430ml (milliliters).}
    \item{The average volume of a piece of candy is 2ml (milliliters).}
\end{itemize}

 \begin{figure}[H] 
 \centering
 \includegraphics[width=.35\linewidth]{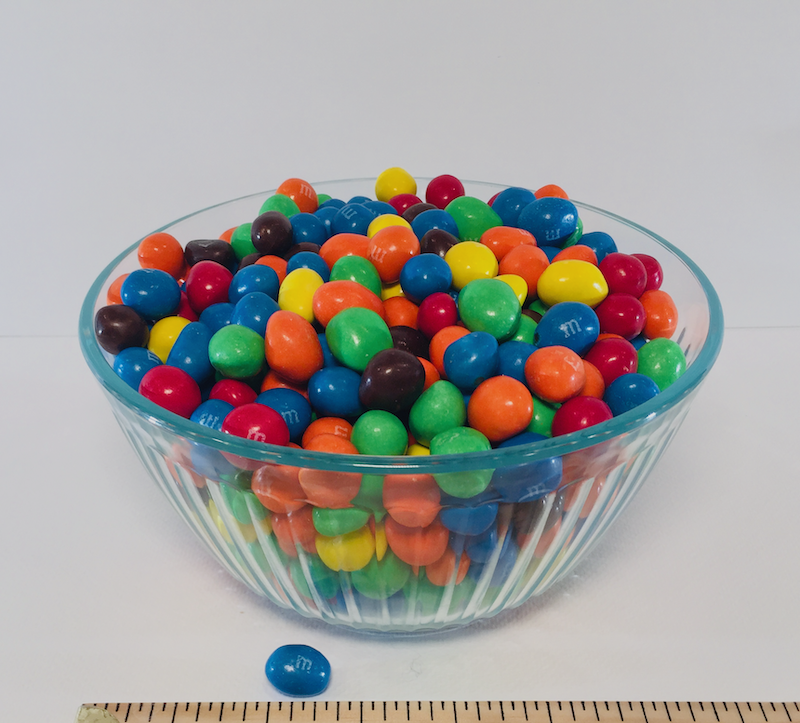}
 \caption{The image for the scale aid: The associated text says ``Below is a picture of the same bowl as before, but with a ruler next it and an additional candy next to the ruler. What is your estimate of the number of candies in the bowl?''}
 \label{fig:ruler_candies}
 \end{figure}

 \begin{figure}[H] 
 \centering
 \includegraphics[width=.5\linewidth]{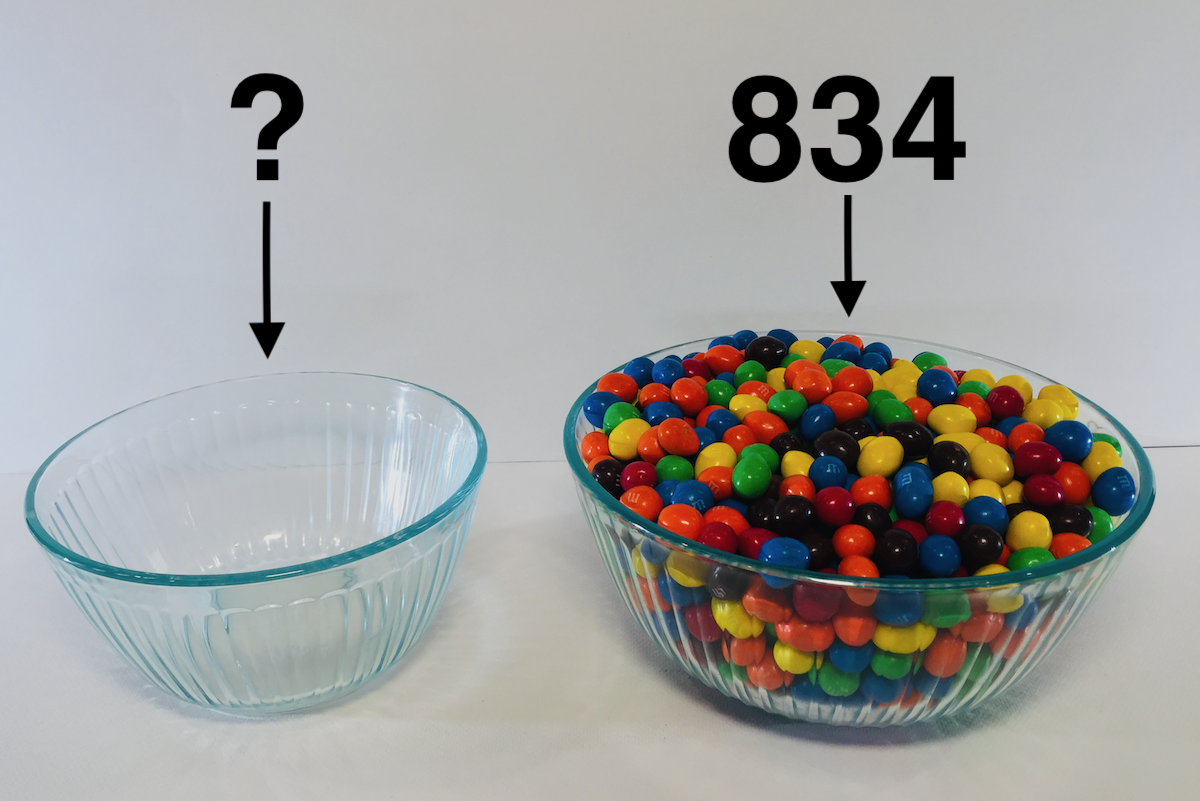}
 \caption{The image for the comparison aid: The associated text says ``The same bowl as above is emptied and shown on the left hand side of this picture. On the right hand side is a larger bowl with 834 candies of the same type in it. What is your estimate of the number of candies in the bowl in the top picture?''}
 \label{fig:comp_candies}
 \end{figure}
The scale and comparison aids were straightforward to create. To create the equation aid, we measured the volume of water necessary to fill the bowl to the brim three times. The three measurements varied less
than a milliliter. To assess the average volume of both types of candies, we found the volume of the water
displaced by placing 50 candies in a measuring cup filled to the brim. This was done with three different
sets of 50 candies previously in the bowl in the pictures. Each final displacement measurement was within
0.5mL. That volume divided by 50 is the average volume of the candies provided to the participants. This procedure was used for both types of candies.

 \begin{figure}[H] 
 \centering
 \includegraphics[width=.5\linewidth]{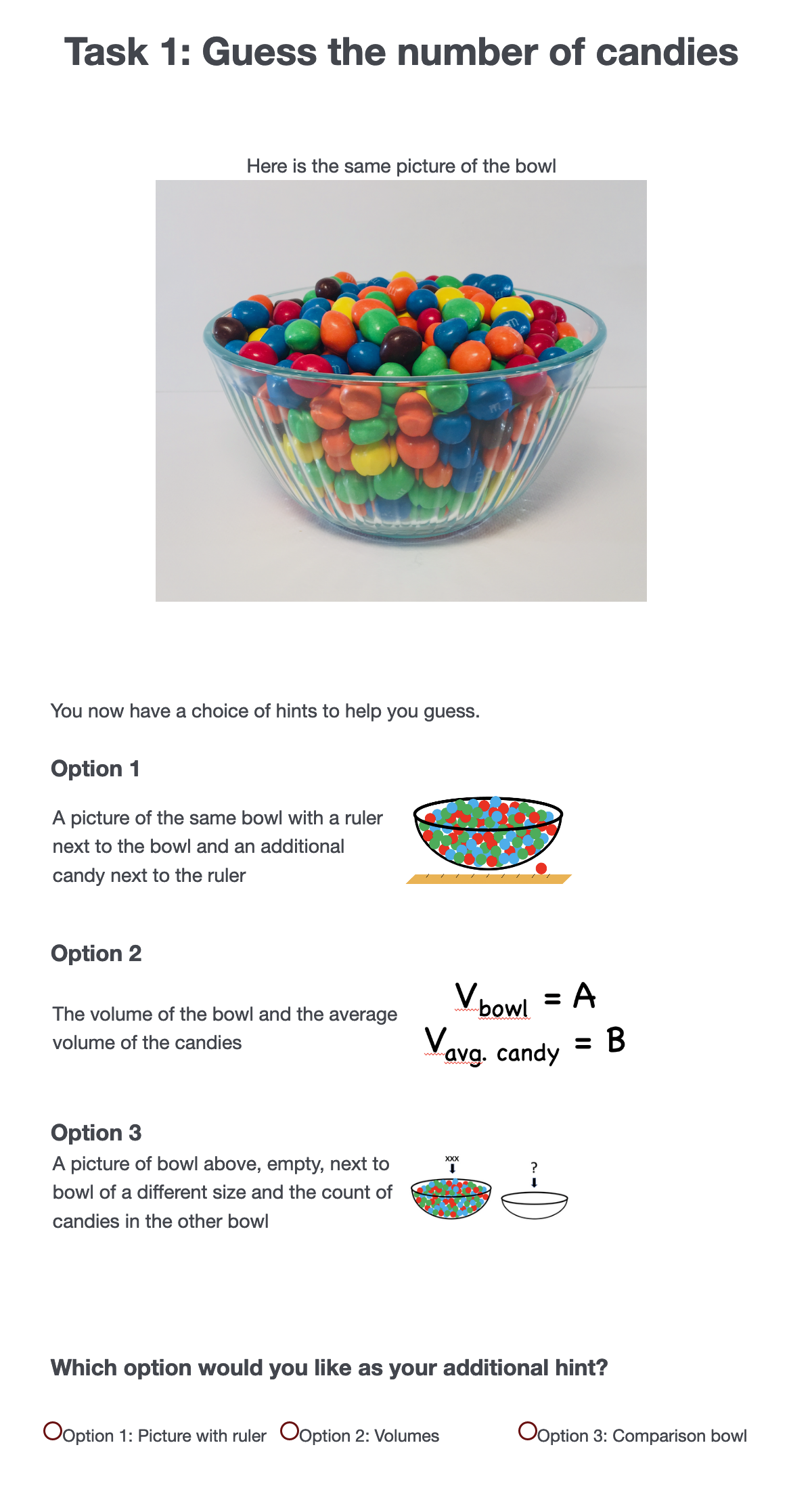}
 \caption{The aid choice interface for the Choice and Multiple Choice conditions. The presentation of options was counterbalanced in our experiments. Participants in the Multiple Choice condition had the option to revisit the choice interface any number of times they desired. We observed no evidence of an effect of the ordering of the choice menu on the choices.}
 \label{fig:Choice prompt}
 \end{figure}

The study was conducted in accordance with a protocol approved by Stanford University IRB and University of Southern California IRB.

\subsection{Study 2 details} \label{sm: study2}
Like Study 1, Study 2 also collected data for a \textit{Multiple Choice} condition for all three collections.  The results for this condition appear in Table \ref{tab:multi_CPI} and do not different substantially from those for the Choice condition. This is likely due to the relatively low use of multiple aids (Nov. '21: 220, 14, 64; May '23: 239, 11, 49; Aug '23: 230, 18, 49, for one, two and three aids respectively).

\begin{table}[h!]
\caption{CPI Forecasting Experiment: Results for the Multiple Choice condition. The reported CPI values appear under the relevant month in the leftmost column.} GSE is the group squared error. MSE is the mean squared error.
\label{tab:multi_CPI}
\centering
\small
\renewcommand{\arraystretch}{1.2}
\begin{tabular}[]{ @{\extracolsep{.1em} } c | c || c | c | c | c | c | c | c }
\shortstack{CPI Forecast\\Date and Value} & \shortstack{Choice\\Treatment} & \shortstack{Decision Aid\\Treatment} & N & \shortstack{Mean\\Estimate} & GSE & MSE \\
\hline
\hline
\multirow{4}{*}{\shortstack{November 2021\\$\text{CPI} = 6.8$}}
& \multirow{4}{*}{Multiple Choice} & Predictive Model & 182 & 7.41 & 0.37 & 5.64 \\
& & Fed Statement & 56 & 6.65 & 0.02 & 1.46 \\
& & Components & 59 & 10.12 & 11.04 & 58.65 \\
\cline{3-7}
& & \textit{All} & 297 & 7.8 & 1.01 & 15.38 \\

\hline
\hline
\multirow{4}{*}{\shortstack{May 2023\\$\text{CPI} = 4.0$}}
& \multirow{4}{*}{Multiple Choice} & Predictive Model & 149 & 5.07 & 1.15 & 1.78 \\
& & Fed Statement & 60 & 5.46 & 2.12 & 8.14 \\
& & Components & 90 & 7.24 & 10.49 & 50.36 \\
\cline{3-7}
& & \textit{All} & 299 & 5.8 & 3.24 & 17.68 \\

\hline
\hline
\multirow{4}{*}{\shortstack{August 2023\\$\text{CPI} = 3.7$}}
& \multirow{4}{*}{Multiple Choice} & 
Predictive Model & 152 & 3.64 & 0.0 & 7.98 \\
& & Fed Statement & 85 & 3.68 & 0.00 & 30.14 \\
& & Components & 85 & 6.76 & 9.37 & 60.92 \\
\cline{3-7}
& & \textit{All} & 298 & 4.84 & 1.29 & 46.1 \\

\hline
\end{tabular}
\end{table}
Our design also included three questions after the prediction was submitted. The first asked ``To the best of your knowledge, what was the reported CPI value for last month?" The reported value here is weakly correlated with accuracy on the primary prediction task. We also asked ``How knowledgeable about national economic trends do you consider yourself?'' (1-Not at all, 7-Extremely) and ''How active are you in managing financial investments (for example, stocks, bonds, or cryptocurrencies)? If you don't currently have any investments, please select not applicable.'' (1-Not at all, 7-Extremely). These were included to test whether these activities are correlated with accuracy on the primary task, which they are not.

\subsubsection{Study 2 materials}
Below are the materials for the November 2021 collection. The materials for the May 2023 and the August 2023 collections have the same design, although the content is updated. The Predictive Model in the later collections was constructed using the same parameters and processes. The materials themselves are available in the Data \& Code supplement.

 \begin{figure}[H] 
 \centering
 \includegraphics[width=.5\linewidth]{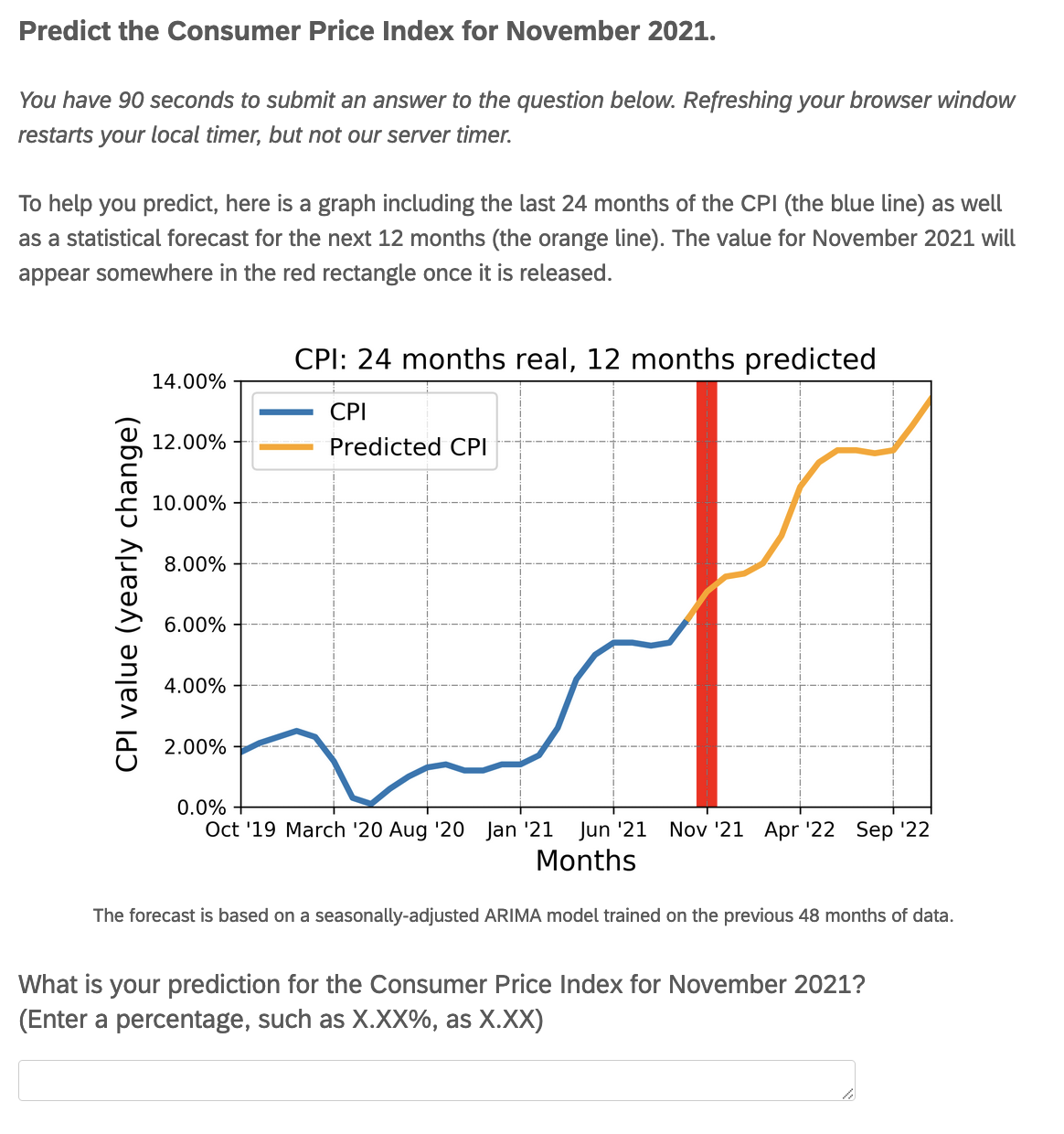}
 \caption{The Predictive Model aid: The details of how the model was calibrated is available in the Code and Data supplement.} A submission timer of 90 seconds was visible.
 \label{fig:arima}
 \end{figure}

 \begin{figure}[H] 
 \centering
 \includegraphics[width=.5\linewidth]{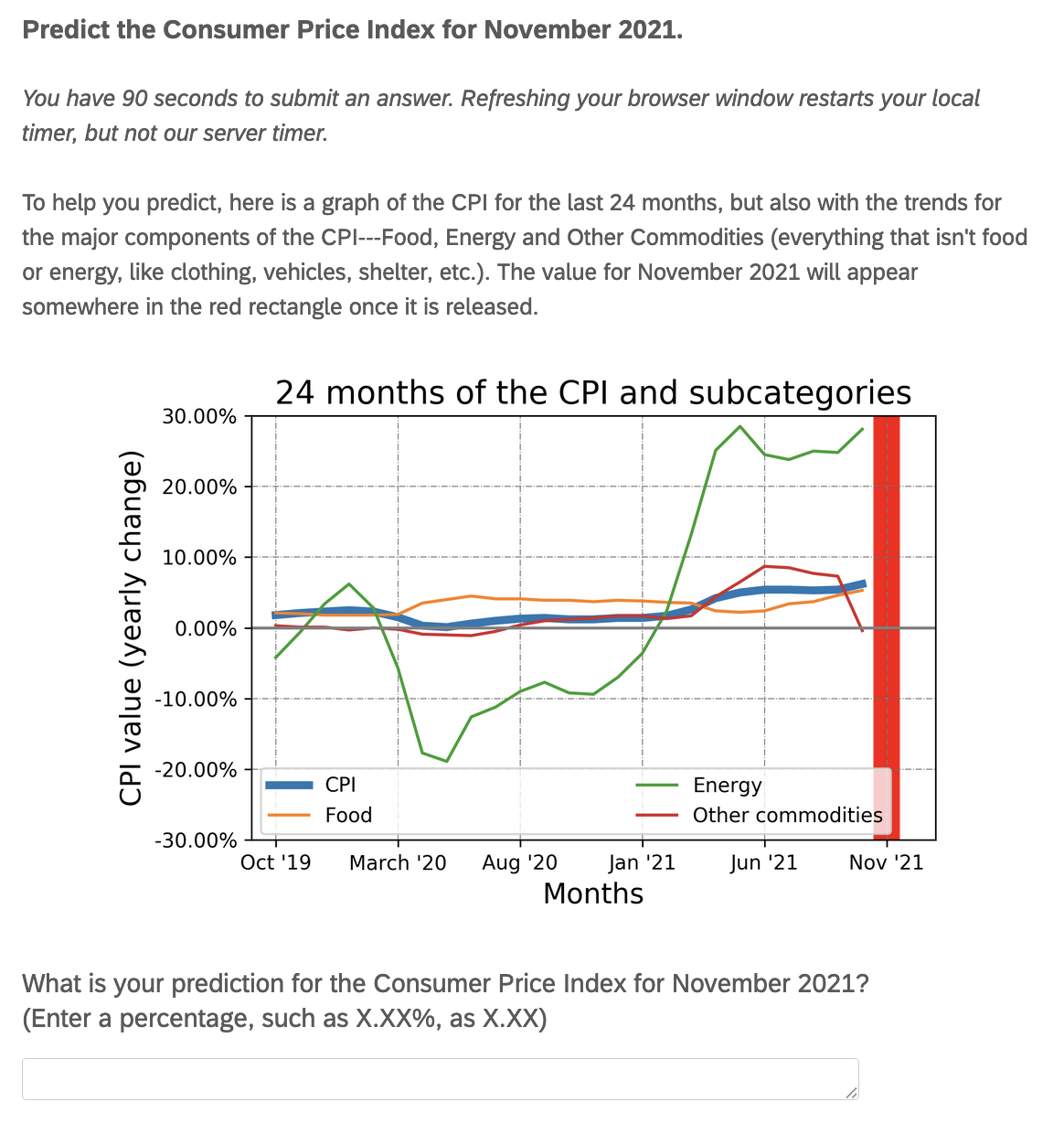}
 \caption{The Components aid: The values of the three main components are a standard part of the Bureau of Labor Statistics' month report and are drawn from the BLS website.}  A submission timer of 90 seconds was visible.
 \label{fig:component}
 \end{figure}

 \begin{figure}[H] 
 \centering
 \includegraphics[width=.5\linewidth]{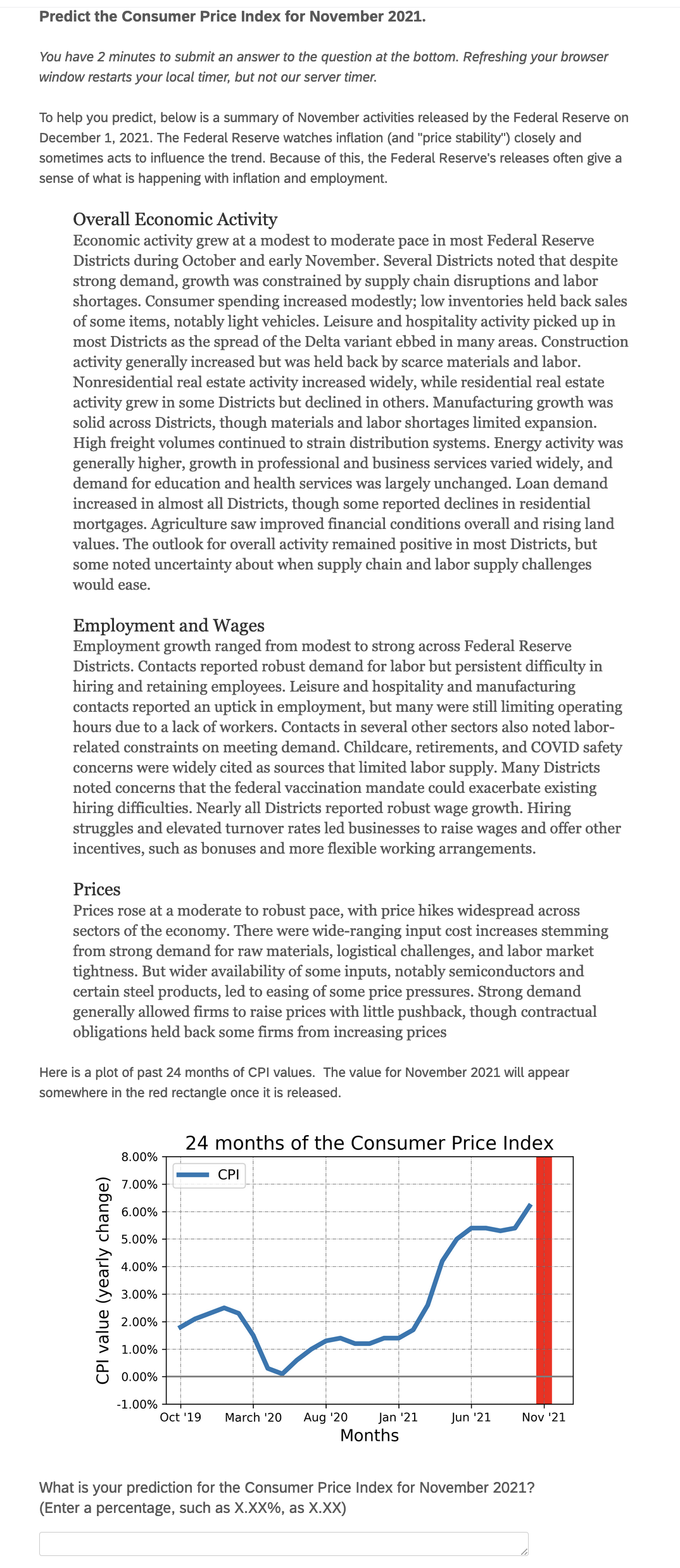}
 \caption{The Federal Reserve statement aid: The statement itself was an image to prevent easy copying and pasting  in order to conduct searches for additional information in Federal Reserve materials.}  A submission timer of 120 seconds was visible.
 \label{fig:choice_prompt}
 \end{figure}

 \begin{figure}[H] 
 \centering
 \includegraphics[width=.5\linewidth]{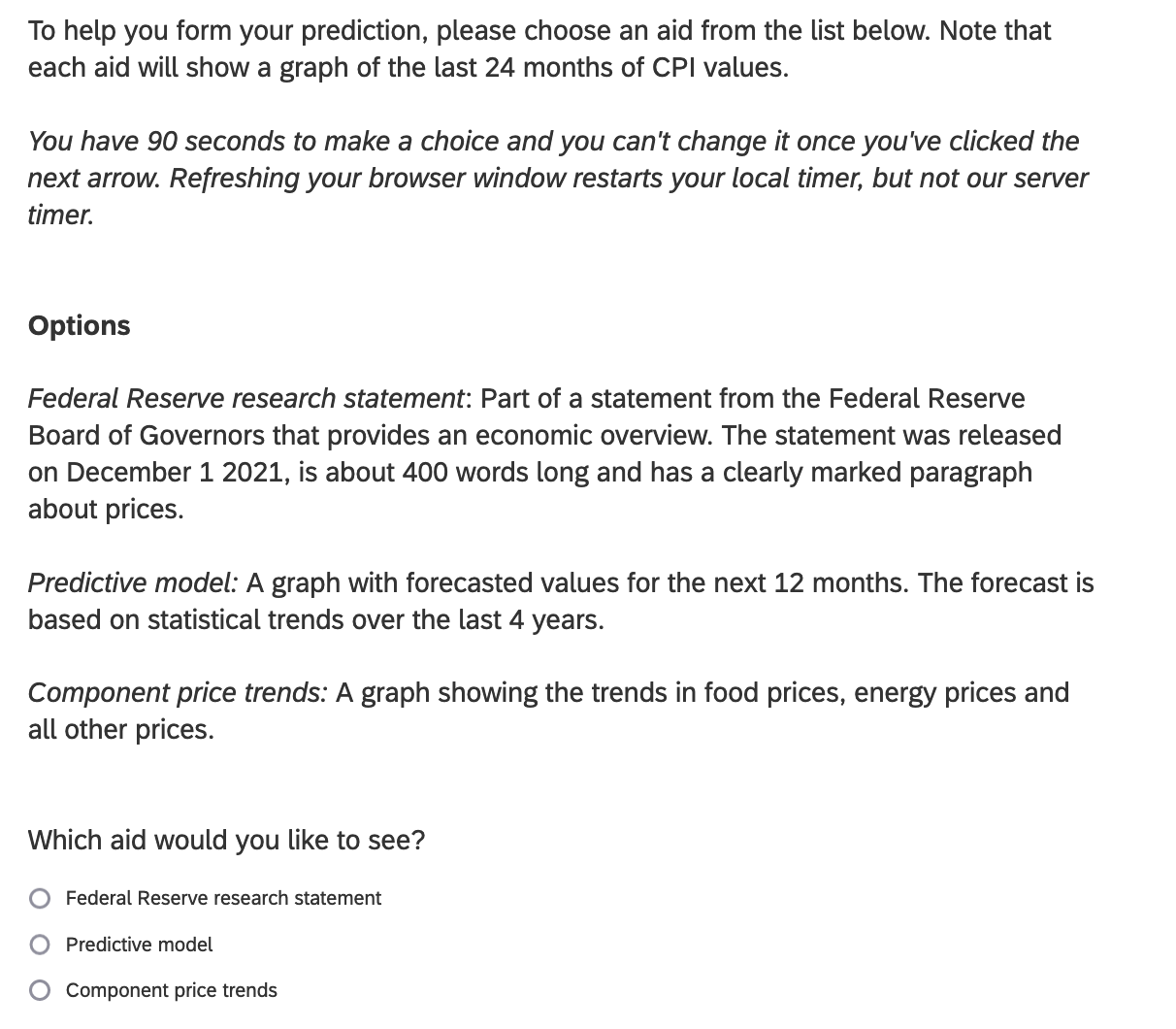}
 \caption{Example of the aid choice prompt: There were six different versions to address the possibility of ordering effects in the appearance of individual aids in the choice menu. There was no evidence of ordering effects. }
 \label{fig:choice_prompt_candy}
 \end{figure}
The study was conducted in accordance with a protocol approved by Stanford University IRB and University of Southern California IRB.

\subsection{Study 3 details} \label{sm: study3}
This study differs from the previous in that it elicits a distribution of values as the estimate.
Because the interface will be novel to participants, we demonstrate it with a practice round before having them proceed to an initial, un-aided round and then finally, an aided round.

The procedure for eliciting a distributional estimate is as follows: participants are shown the image of objects in a glass container (hard candies for the practice round and nickels for the primary rounds) and asked to submit their best guess for the number of objects. They are then asked to provide a lowest and highest ``number of [objects] you think might be possible.'' Inputs are validated to be non-negative and lower/higher than the best guess.  These inputs are used to initialize an interactive normal distribution plot on the next page. The best guess is the mean of the distribution and is visualized as a point. The larger of the two absolute differences between the mean and lowest and highest estimates becomes the value at the 99.9th percentile (e.g., 3 standard deviations from the mean). From the 99.9th percentile, we impute and visualize the standard deviation for the distribution as two points on either side of the mean. The lowest and highest estimates are also visualized on the plot. The mean and standard deviation points can be clicked and dragged to shape the distribution. While the initial low and high parameters did not need to be symmetrical with respect to the best guess, the interactive plot does enforce a symmetrical distribution. The axes are dynamically updated so that distribution occupies most of the plotting area. 

The interactive page shows the primary image and decision aid as before. It also explains that accuracy bonuses are determined by density of their distribution of estimates at the true value. The maximum value of the bonus is \$50 and can only occur when the mean is the true value and standard deviation is less than 1 unit. The logic of this scheme is likely not intuitive to participants, so the interactive plot allows them to move around a hypothetical true value point on the x-axis. From that point, there is a vertical line to the curve representing the distribution. From that point of intersection, a horizontal line runs to the y-axis, where the scale is value of the bonus. The current best guess, the hypothetical true value, and the hypothetical bonus amount for the hypothetical true value are all displayed as text to the right of the interactive plot. There is also a button allowing them to reset the graph (and their estimates) back to their initial inputs. Participants have an unlimited amount of time to refine their estimates using the interactive plot before submitting. 

For the practice round only, the participants are then shown the true value and the bonus they would have gotten given their submitted distribution. These results are displayed as text and on the same style plot as before, but without the interactive elements. The primary goal of the practice round is to help participants understand what their estimate actually is (i.e., a distribution) and to clarify how it is linked to accuracy bonuses.

Participants repeat this procedure for the nickels-in-a-cylinder prediction task without knowing that they will also be given the opportunity to do it again with the benefit of a decision-making aid. Once they have completed this round, they are told they are going to be given a ``hint'' that might help them improve their estimate. They are asked to rank three ``hints'' in the order they prefer them. The same three types of aids as used for Study 1 are used here (\textit{Equation}, \textit{Comparison}, and \textit{Scale}). We display stylized images of the aids on tile elements. Hovering above the tile reveals a short textual description of the visualized aid. 

To indicate their preference for aids, participants move the tiles into a vertical stack. The highest position is labeled has having a rank of 1 and a probability of being shown as 70\%. The lowest position, rank 3, has a probability of 10\%. Participants can adjust their ranking until they submit. 

An aid is randomly selected given the probabilities of the participant's preference ordering and displayed alongside the original nickels-in-a-cylinder image. They are asked to submit their best guess and the low and high range. They then are taking to the interactive distribution where they can refine their guess. The original image and the hint are available on this page as well. 

\subsubsection{Study 3 materials}
The interface for this study was developed using the Empirica platform \citep{almaatouq_empirica_2021}.

 \begin{figure}[H] 
 \centering
 \includegraphics[width=.65\linewidth]{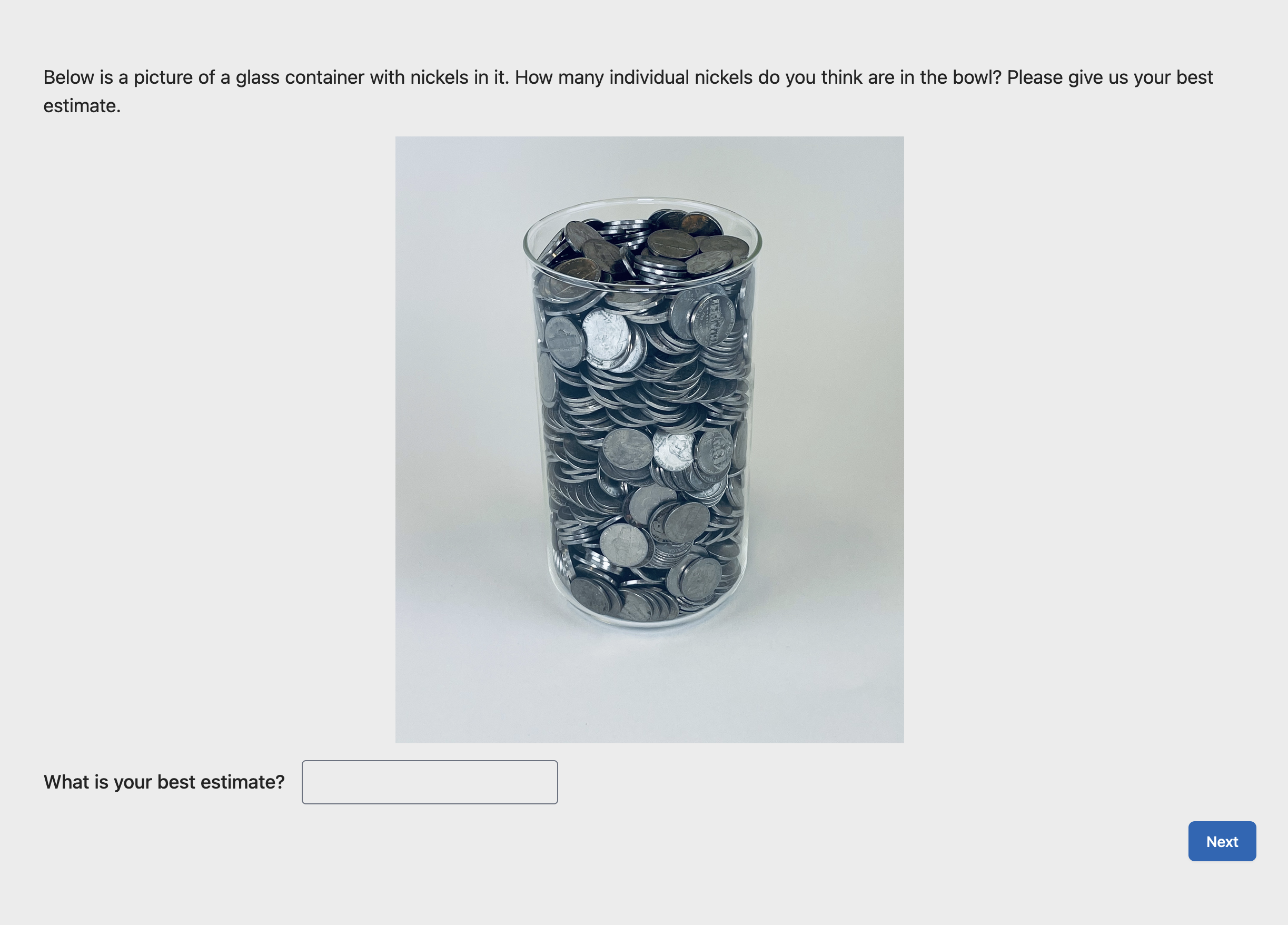}
 \caption{Initial estimate prompt, best estimate.}
 \end{figure}
 \begin{figure}[H] 
 \centering
 \includegraphics[width=.65\linewidth]{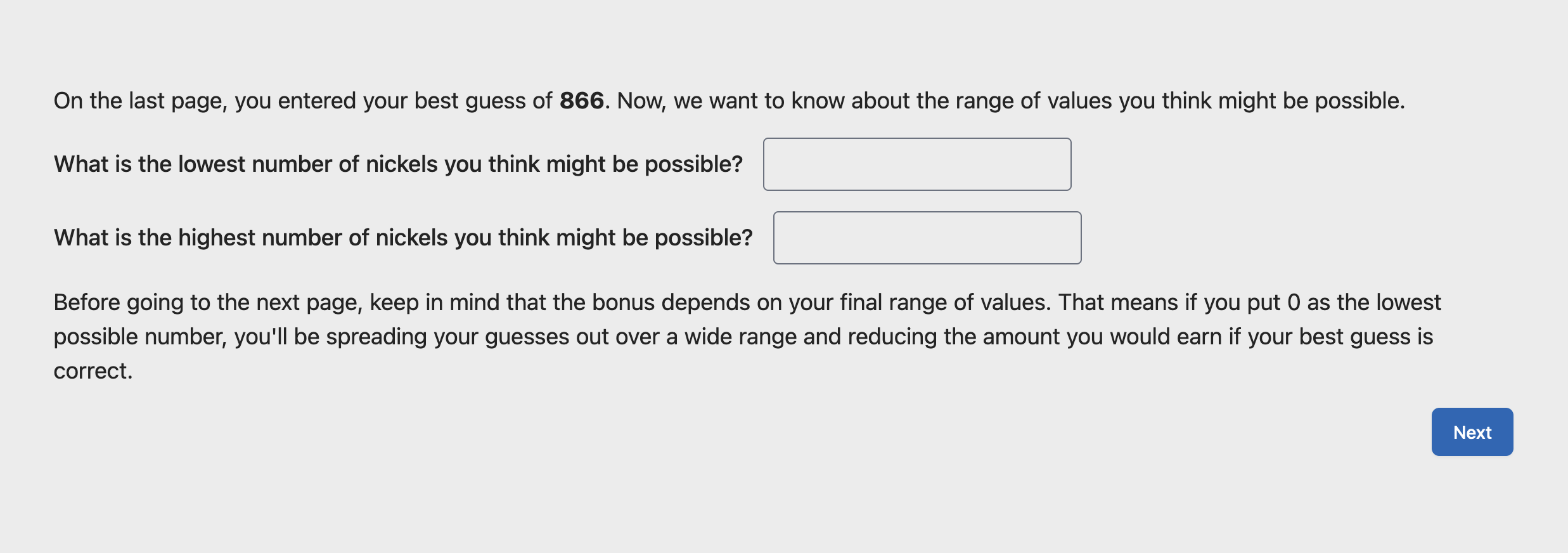}
 \caption{Initial estimate prompt, high and low extrema.}
 \end{figure}
 \begin{figure}[H] 
 \centering
 \includegraphics[width=.65\linewidth]{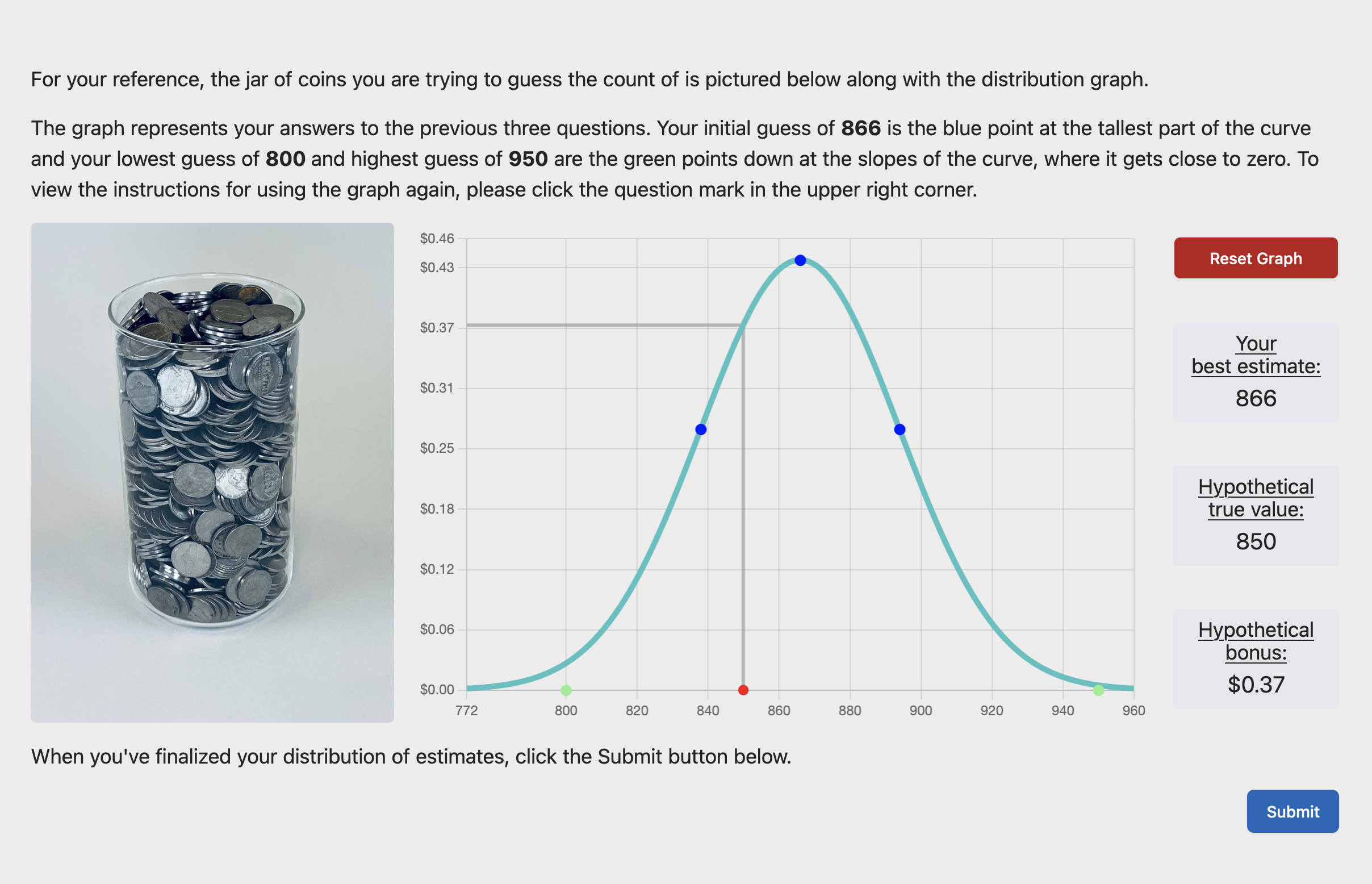}
 \caption{Initial estimate prompt, interactive distribution with hypothetical true value moved to show the corresponding hypothetical bonus.}
 \end{figure}
 \begin{figure}[H] 
 \centering
 \includegraphics[width=.65\linewidth]{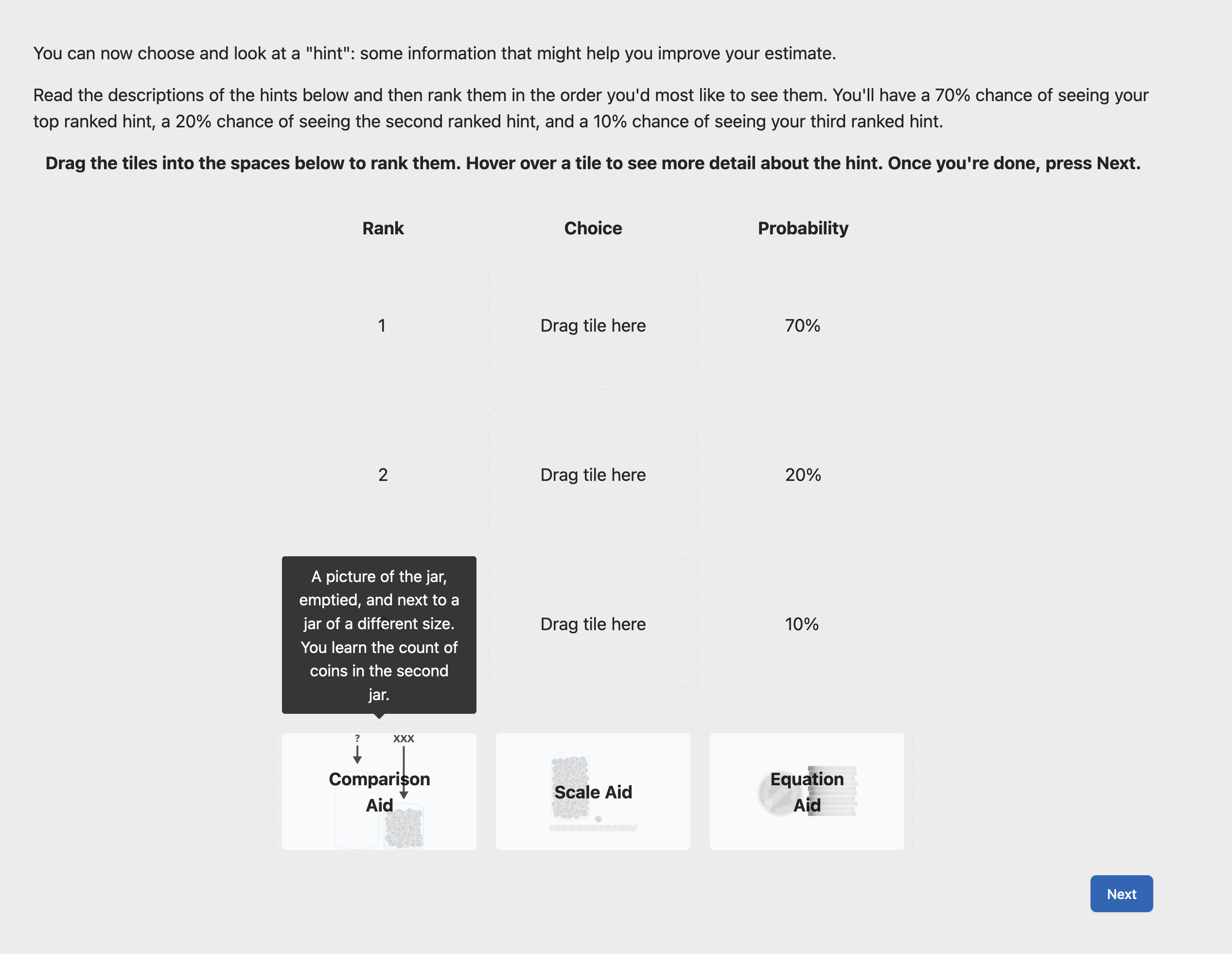}
 \caption{Aid choice prompt (before selections): Each decision aid ``tile'' provided a text description when the cursor hovered over it. The equation aid description read ``The volume of the jar and the average volume of 10 coins''. The scale aid description read ``A picture of the jar with a ruler next to it and an additional coin next to the ruler.'' The comparison aid description read ``A picture of the jar, emptied, and next to a jar of a different size. You learn the count of coins in the second jar.}
 \end{figure}

 \begin{figure}[H] 
 \centering
 \includegraphics[width=.65\linewidth]{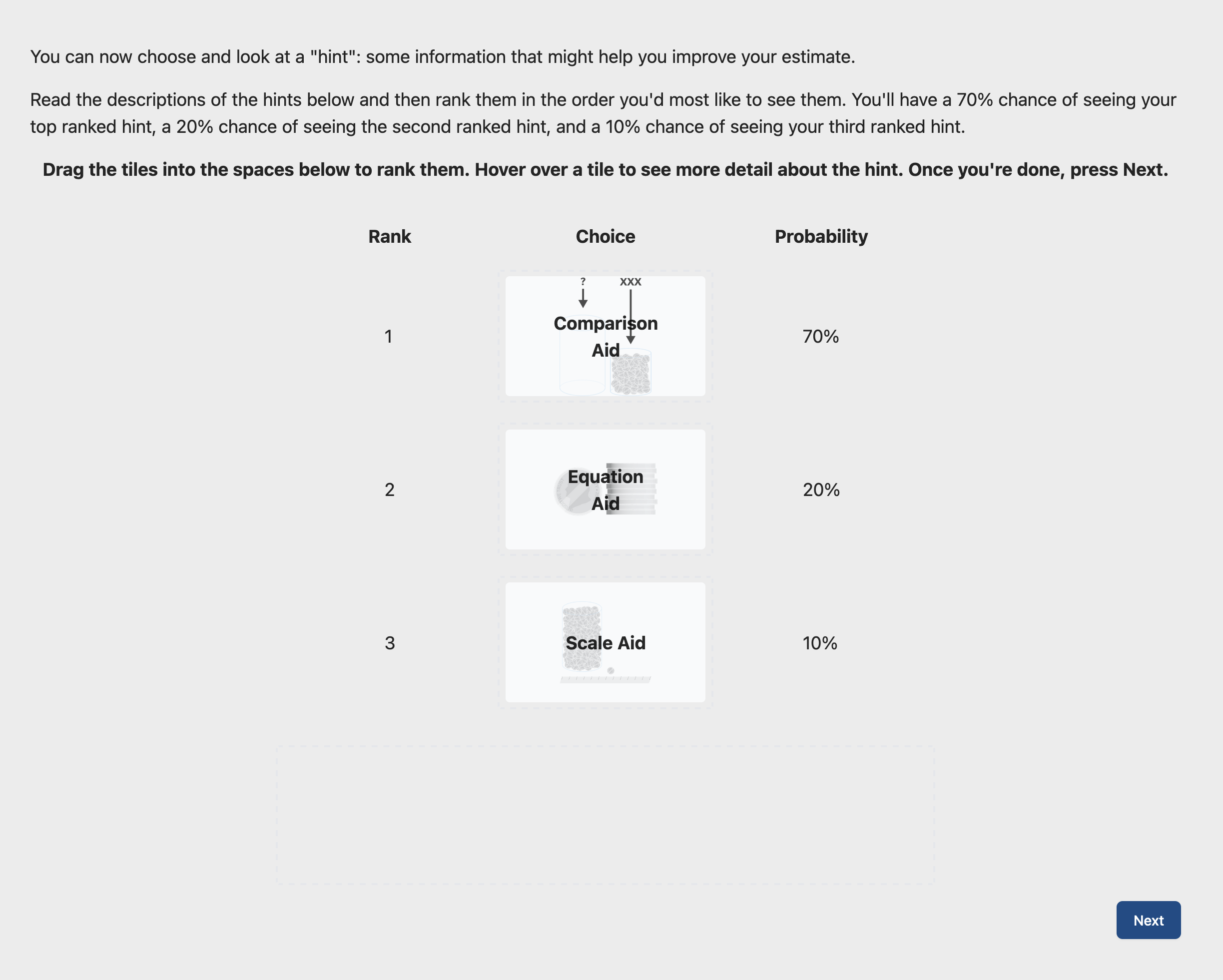}
 \caption{Aid choice prompt, with example selection.}
 \end{figure}
 \begin{figure}[H] 
 \centering
 \includegraphics[width=.65\linewidth]{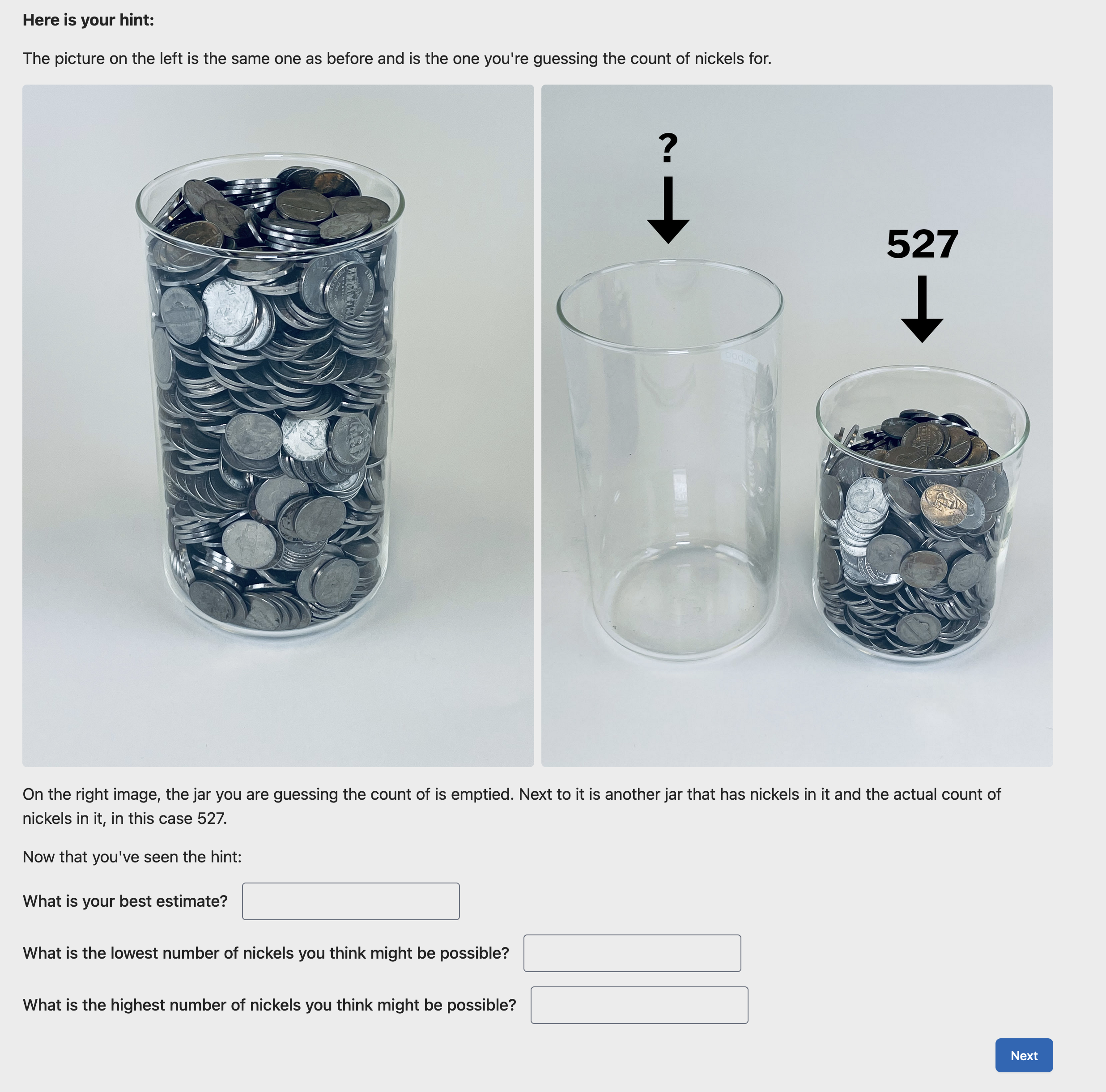}
 \caption{Main prompt, with comparison aid.}
 \end{figure}
 \begin{figure}[H] 
 \centering
 \includegraphics[width=.65\linewidth]{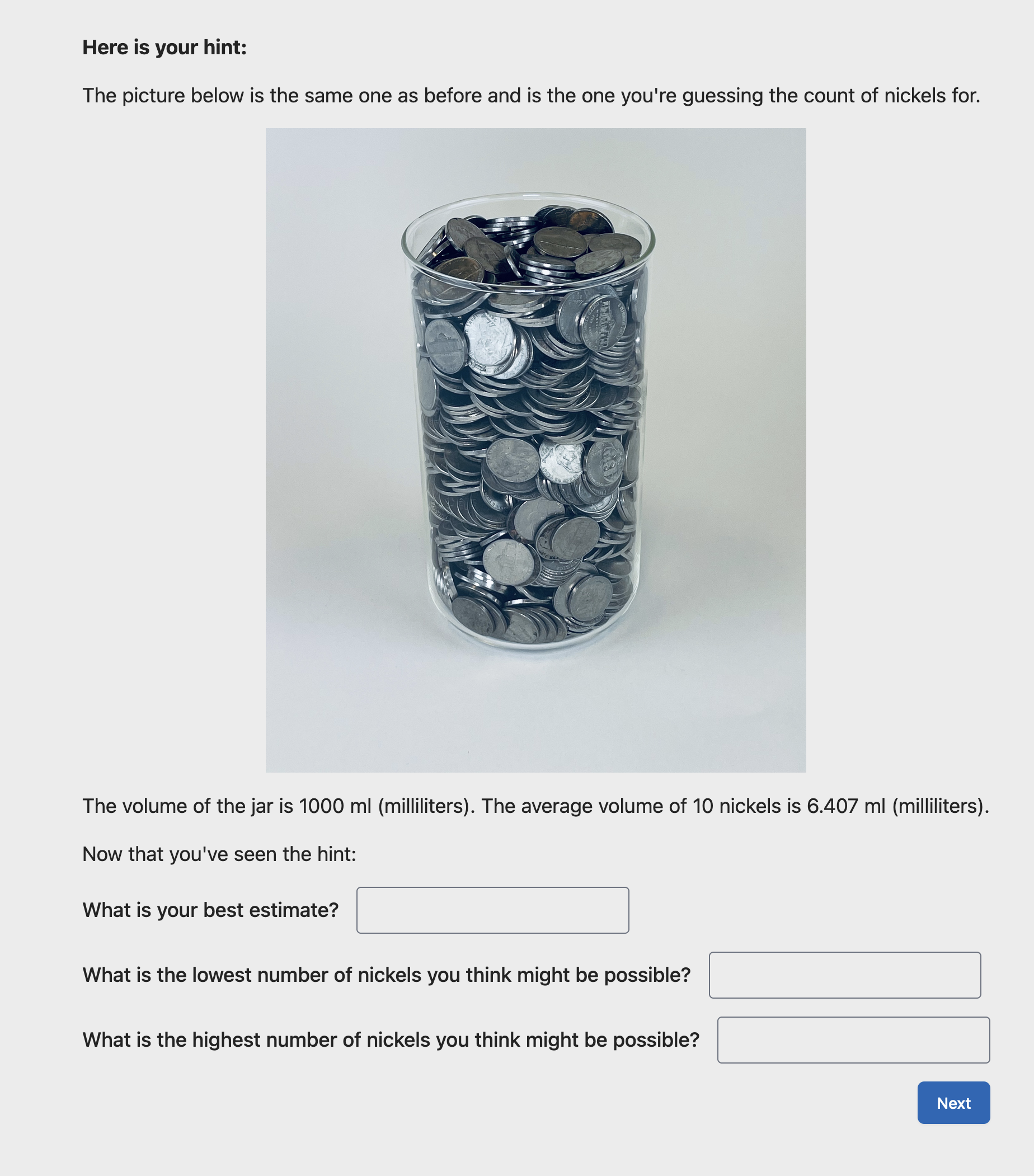}
 \caption{Main prompt, with equation aid}
 \end{figure}
 \begin{figure}[H] 
 \centering
 \includegraphics[width=.65\linewidth]{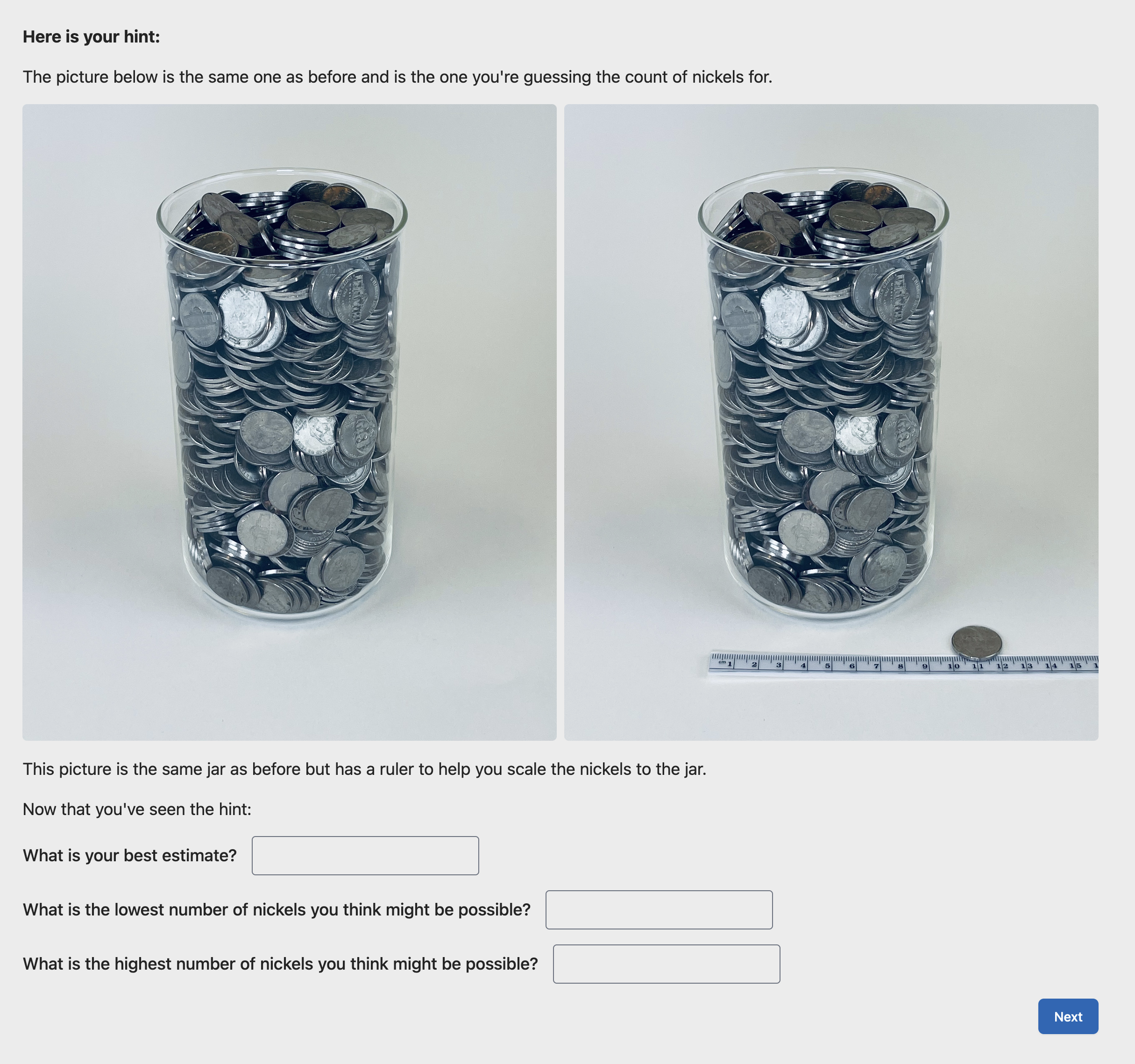}
 \caption{Main prompt, with scale aid.}
 \end{figure}
 \begin{figure}[H] 
 \centering
 \includegraphics[width=.65\linewidth]{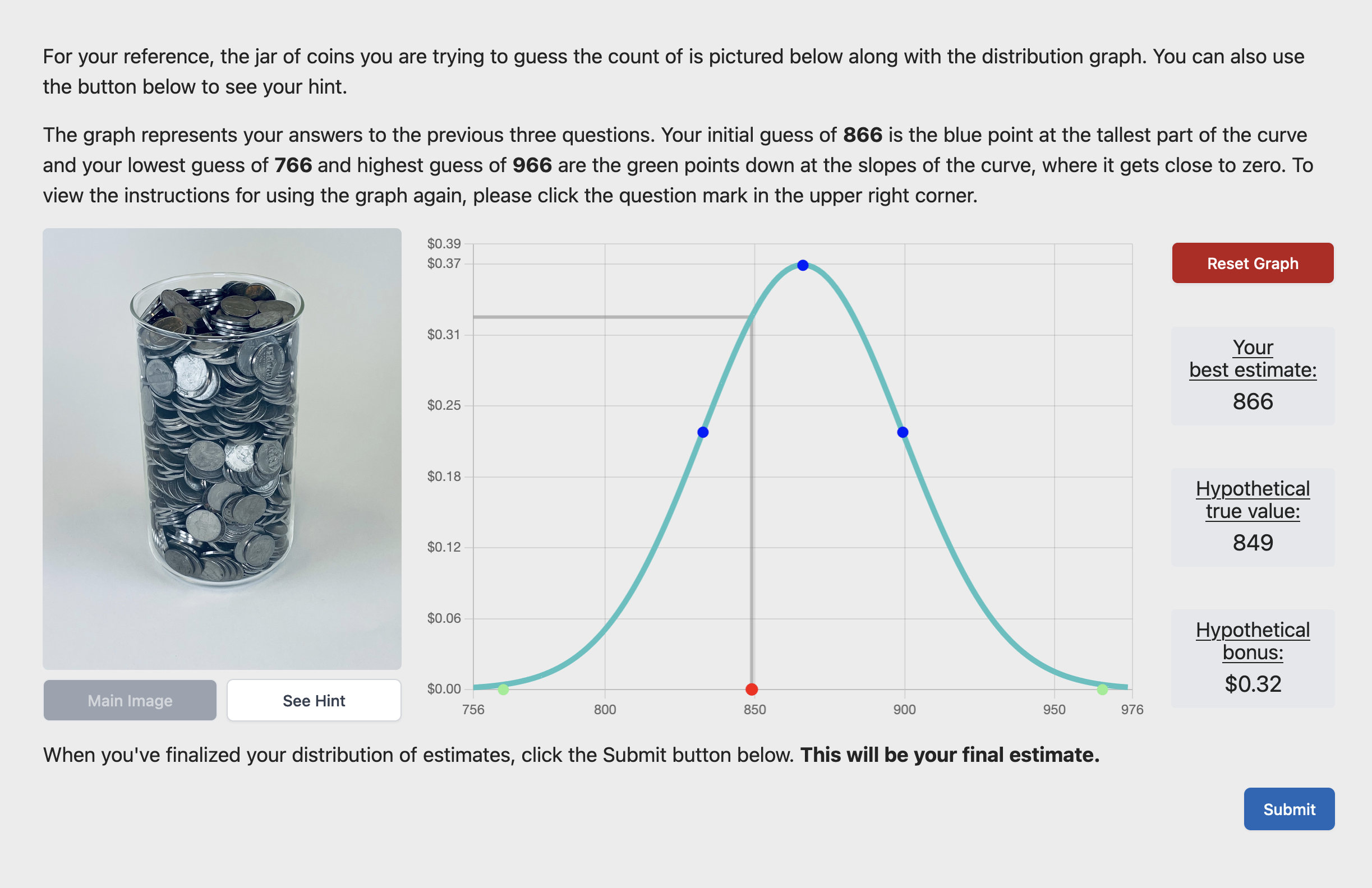}
 \caption{Main prompt: Interactive distribution (view 1), with the hypothetical true value altered to show the corresponding hypothetical bonus. Main image visible.}
 \end{figure}
 \begin{figure}[H] 
 \centering
 \includegraphics[width=.65\linewidth]{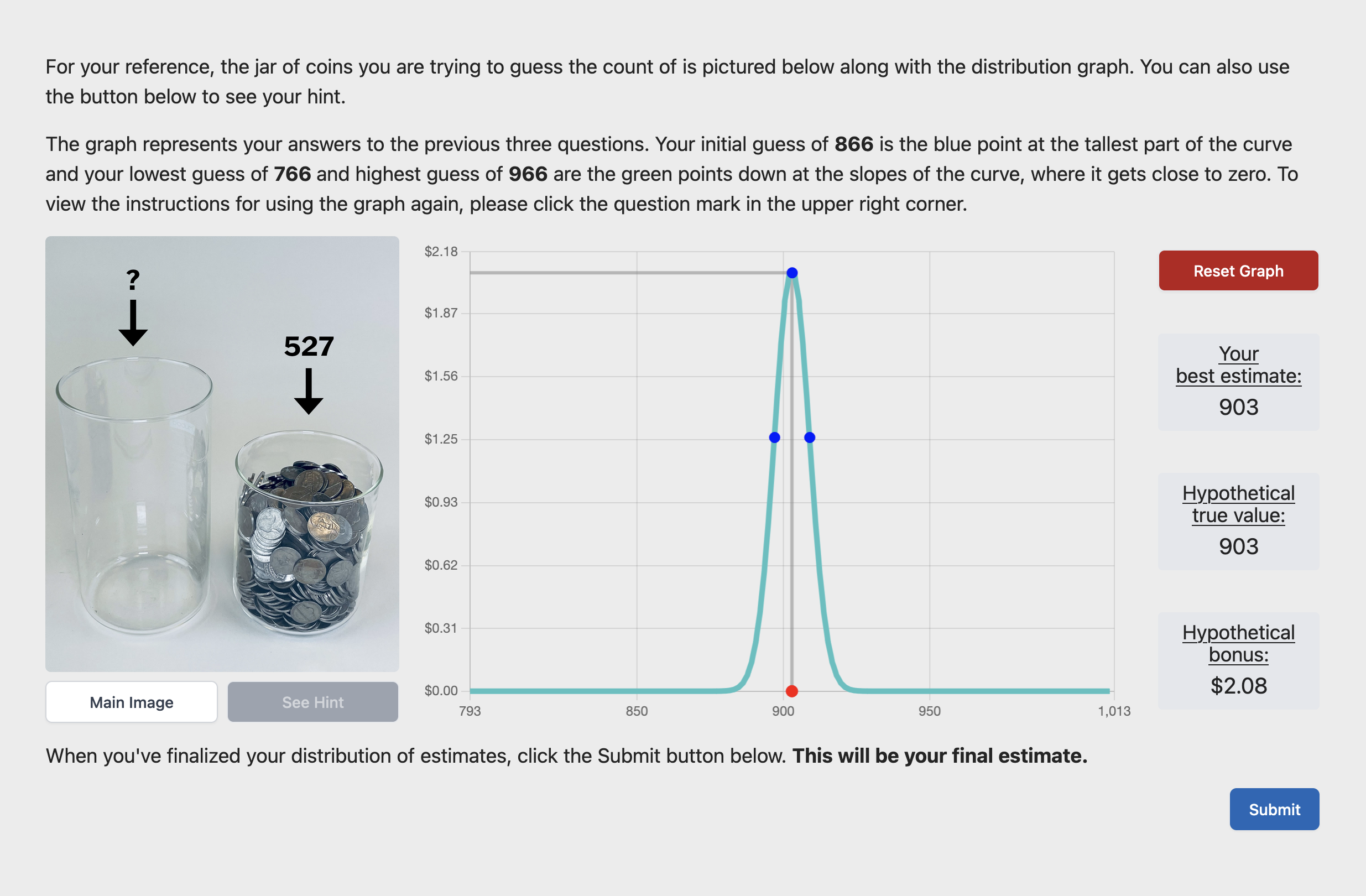}
 \caption{Main prompt: Interactive distribution  (view 2) with altered distribution and ``hint'' visible.}
 \end{figure}
The study was conducted in accordance with a protocol approved by Stanford University IRB.
\end{document}